\documentclass[aps,floatfix,preprint,preprintnumbers,nofootinbib,superscriptaddress,natbib]{revtex4}
\usepackage{siunitx}
\usepackage[titletoc]{appendix}
\usepackage{graphicx,float}
\usepackage{epsfig}		
\usepackage{dcolumn}
\usepackage{bm}
\usepackage{subfigure}
\usepackage{lipsum}
\usepackage{fancyhdr}
\usepackage[all]{xy}
\usepackage{amsmath,upgreek}
\usepackage{amssymb}
\usepackage{empheq}
\usepackage{xcolor}
\usepackage{bbold}
\usepackage{comment}
\definecolor{lightgreen}{HTML}{90EE90}

\usepackage{pdfpages}
\usepackage{color}
\usepackage{graphicx,epstopdf}
\usepackage[colorlinks=true,
linkcolor=red,
urlcolor=purple,
citecolor=blue,hyperindex]{hyperref}
\usepackage{cleveref}
\def\0{\mbox{\tiny $0$}}
\def\1{\mbox{\tiny $1$}}
\def\2{\mbox{\tiny $2$}}
\def\3{\mbox{\tiny $3$}}
\def\4{\mbox{\tiny $4$}}
\def\5{\mbox{\tiny $5$}}
\def\6{\mbox{\tiny $6$}}
\def\7{\mbox{\tiny $7$}}
\def\8{\mbox{\tiny $8$}}
\def\9{\mbox{\tiny $9$}}

\def\f14{\mbox{\tiny $\frac{1}{4}$}}

\DeclareMathOperator{\Tr}{\mbox{Tr}}
\DeclareMathOperator{\Res}{\mbox{Res}}

\makeatletter\def\Hy@Warning#1{}\makeatother
\begin{document}
	\title{Intrinsic correlations for statistical ensembles of Dirac-like structures}
	\author{Caio Fernando e Silva}
	\email{caiofernandosilva@df.ufscar.br}
	\author{Alex E. Bernardini}
	\email{alexeb@ufscar.br}
	\affiliation{~Departamento de F\'{\i}sica, Universidade Federal de S\~ao Carlos, PO Box 676, 13565-905, S\~ao Carlos, SP, Brasil.}
	
	\date{\today}
\begin{abstract}
The Weyl-Wigner formalism for evaluating the intrinsic information of Dirac bispinors as correlated qubits (localized) in a magnetic field is investigated in the extension to statistical ensembles. The confining external field quantizes the quantum correlation measures implied by the spin-parity qubit structure of the Dirac equation in 3+1 dimensions, which simplifies the computation of the entanglement quantifier for mixed states in relativistic Landau levels. This allows for the evaluation of quantum and classical correlations in terms of entropy measures for Dirac structures that are eventually mixed. Our results are twofold. First, a family of mixed Gaussian states is obtained in phase space, and its intrinsic correlation structure is computed in closed form. Second, the partition function for the low-dimensional Dirac equation in a magnetic field is derived through complex integration techniques. It describes the low-temperature regime in terms of analytically continued Zeta functions and the high temperature limit as a polynomial on the temperature variable. The connection with lower dimensional systems is further elicited by mapping the spin-parity qubits to valley-sublattice bispinors of the low-energy effective Hamiltonian of graphene.
\end{abstract}

\pacs{}
\keywords{Dirac Spinors, Statistical Ensembles, Intrinsic Correlations, Partition Function}
\date{\today}
\maketitle

\section{Introduction}

In spite of being proposed through distinct quantum mechanical platforms, information processing with statistical ensembles is still a challenging task in the context of its continuously investigated open theoretical issues. Discrete and continuous quantum information issues are indeed identified in numerous arrangements \cite{Andersen2}, from qubit systems \cite{n024,Vedral,Henderson} to high dimensional states \cite{Weedbrook}. Of course, they do not exhaust all the possibilities, since incremental degrees of freedom, for instance, related to internal symmetries, can also be necessary for encompassing the quantum information content of a physical system. That is the case of relativistic-like systems which have also been shown to accommodate a consistent information approach to spinors \cite{extfields,n010,BernardiniEPJP}. In particular, Dirac-like structures have been described in a plethora of physical systems, standing as platforms for implementing quantum information protocols \cite{n001, n002, n004, n005, n006, graph03, graph04,CastroNeto,MeuPRB,PRBPRB18,New} where the intrinsic entanglement -- related to internal degrees of freedom -- plays a determinant role.

As a matter of fact, the quantum information profile of Dirac bispinors has been thoroughly investigated with distinct types of interacting potentials, as supported by the $SU(2)\otimes SU(2)$ group structure implied by the Dirac equation \cite{MeuPRA}. In other words, two discrete degrees of freedom, namely, spin and parity, have their entanglement affected by external fields, through interactions supported by the Poincar{\'e} group of symmetry \cite{Thaller,extfields}. In addition, such an associated internal correlation structure is also modified by phase-space variables when the Hamiltonian has some explicit dependence on position variables \cite{BernardiniEPJP,Bermudez,Rusin}. To summarize, Dirac bispinors can be regarded as two-qubit entangled structures \cite{extfields,n010}, eventually correlated by phase-space variables \cite{MeuPRA}. Thus, the inclusion of position-dependent interactions can be regarded as a natural extension from a qubit theory to a continuous variable approach in Dirac-like structures which, by the way, is relevant for describing Hamiltonian dynamics with a reduced symmetry \cite{Greiner}. 

In this context, despite the subtleties of each physical system, there are also some transversal questions which must be answered to complete understand the nature of correlations in mixed states. The difficulty lies not exclusively on the existence of multiple quantifiers of entanglement, but also on quantifying other types of quantum correlations that are distinct from entanglement \cite{Bera,Olivier,Luo}. These obstacles need to be addressed in order to characterize a physical system as a candidate for quantum computation protocols. Therefore, distinguishing quantum from classical correlations becomes a crucial endeavor.

From our recent results, the Dirac Hamiltonian in a confining magnetic field has been considered in the dynamics of pure Gaussian states, for which the phase and configuration space representations have been shown to be completely equivalent \cite{PRA2023}. Even though this two-way correspondence is formally indisputable, the relevant information measures become much clearer in the Weyl-Wigner representation of quantum mechanics \cite{Weickgenannt, Zhuang,1986,1987, Wigner, Case,BernardiniEPJP,PRA2021}, as disseminated from the quantum information framework with light fields \cite{Andersen}. From those results, one confirms that the phase-space representation is a natural extension for addressing the classical limit and properties of mixed states.

Aside from time-dependent phenomena, external interactions are also expected to influence the quantum information profile of spinors in a more realistic description of a quantum information approach to the Dirac equation \cite{extfields}. However, when the exact details of the environment are not known, an option is to resort to statistical ensembles, instead of solving an even more intricate Hamiltonian. Thus, the specification of the mixed state depends on the type of interaction under investigation. The canonical ensemble is a prime example of such interactions, which can be studied under the Weyl-Wigner formalism in order to assess the quantum information profile of intrinsically correlated bispinors at finite temperatures. 

For all these reasons, the main goal of this work is to obtain the information profile of statistical ensembles in relativistic Landau levels, culminating in the partition function for a quantum information theory at finite temperatures. Considering that randomness is a type of classical correlation, it is essential to distinguish it from true quantum resources, i.e. from quantum correlations. In particular, our analysis focuses on the quantum concurrence as a quantifier of entanglement of formation extended to mixed states.

Thus, the paper is organized as follows. In section II, the intrinsic entanglement measure for mixed Dirac bispinors is revisited and the framework for evaluating intrinsic correlations in terms of entropy measures and Wigner functions is recovered. In section III, a family of mixed Gaussian states in relativistic Landau levels is obtained, and the engendered connection between the continuous-variable entropy and localization is discussed. The obtained results are applied for the analysis of a canonical ensemble of the low-dimensional Dirac equation in section IV. In particular, the partition function and intrinsic correlations in the high-temperature limit are all obtained through analytic continuation techniques. Finally, our conclusions are drawn in section V, where the mapping to graphene Hamiltonians is explicitly discussed, and an outlook for the next developments is pointed out.

\section{Mixed state correlations in quantizing fields}
\label{key}

Quantifying quantum and classical correlations of a quantum state is an essential task. Thus, one first motivates the search for a quantum correlation measure in confined systems for statistical ensembles. Then, entropy and purity measures are introduced for Dirac bispinors as correlated qubits.

\subsection{Quantum concurrence for (localized) two-qubit systems}
For pure two-qubit states, the genuine non-separability criteria is given by the entanglement of formation, whereas the extension to mixed states can be defined as the convex-roof extension of the pure-state entanglement. Formally, this is equivalent to finding the mean value of the pure-state entanglement, minimized over all decompositions on pure states \cite{n024},

\begin{equation}
E_{EoF} [\varrho] = \mbox{min}_{\varrho_k} \displaystyle \sum_k q_k E_{vN}[\varrho_k].
\end{equation}
It is related to the quantum concurrence as 
\begin{eqnarray}
E_{EoF} [\varrho] &=& \mathcal{E}\left[ \frac{1 - \sqrt{1 - \mathcal{C}^2[\varrho]}}{2}\right],
\end{eqnarray}
with $\mathcal{E}[\lambda] =- \lambda \log_2 \lambda - (1-\lambda)\log_2 (1-\lambda)$, and where $\mathcal{C}[\varrho]$ is given for two-qubit systems as 
\begin{equation}\label{qubitconcurrence}
\mathcal{C}[\varrho] = \mbox{max}\{ \omega_1 - \omega_2 - \omega_3 - \omega_4 \, , \,0 \},
\end{equation}
where $\omega_1 > \omega_2 > \omega_3 > \omega_4$ are the eigenvalues of the operator $\sqrt{\, \sqrt{\varrho} \, (\sigma_y \otimes \sigma_y) \varrho^\ast (\sigma_y \otimes \sigma_y) \, \sqrt{\varrho}\, }$ and $\varrho^*$ is the complex-conjugated density operator. Thus, quantum concurrence itself is a measure of non-separability. This formula was first derived by finding a set of pure density matrices, with the same concurrence, that realizes $\rho$. In this case, the mean quantum concurrence,
\begin{equation}
\mathcal{C}[\varrho] = \mbox{min}_{\varrho_k} \displaystyle \sum_k q_k \mathcal{C} [\varrho_k],
\label{averagedconcurrence}
\end{equation}
is the same as the pure-state concurrence for the set the minimizes the quantifier, since $\sum_k q_k = 1$. One recalls that the probability distribution $q_k$ is usually not unique. Statistical ensembles are often decomposed over a complete basis in multiple ways. 

For confined systems, however, there is typically a handful or even just one set of pure density matrices that realize a given mixed state. Therefore, Eq.~\eqref{averagedconcurrence} is a valid quantifier in terms of the pure-state entanglement. To verify its relation with other information quantifiers for Dirac bispinors, one first introduces the relevant density matrix for Dirac structures in a localizing external field.
\subsubsection{Qubit structure of the 3+1 Dirac equation}
Before moving to mixed states, however, it is useful to observe that this issue was initially addressed by extending correlation measures to pure localized Dirac bispinors, for which spinor degrees of freedom are regarded as correlated qubits. In this case, the entanglement exhibited by the solutions of the Dirac equation is regarded as intrinsic, or intraparticle, entanglement \cite{MeuPRA}. 

To clear up this assertion, one notices that the qubit structure of the free Dirac equation is obtained in the Dirac representation, for which the gamma matrices are given by $\gamma_{0} = \beta$, $\gamma_{j} = \beta\alpha_j$, $\{\gamma_{\mu},\gamma_{5}\} = 0$, and $\sigma_{\mu\nu} = (i/2)[\gamma_{\mu},\gamma_{\nu}]$. Thus, they can be expressed in terms of Kronecker products of Pauli matrices
\begin{equation}
\label{twoqubithamiltonian}
{H} = \mathbf{k}\cdot ({\sigma}_x^{(P)} \otimes {\bm{\sigma}}^{(S)}) + m ( {\sigma}_z ^{(P)} \otimes {I}_{2}^{(S)}),
\end{equation}
associated to spin $(S)$ and parity $(P)$ degrees of freedom. Dirac bispinors carry the correlation structure associated to these degrees of freedom, which are separable if external fields are absent. In any case, since the momentum is a fixed parameter of the theory, a matrix model is adequate to manage the eigensolutions of this Hamiltonian, emulating the typical formalism in qubit systems. Thus, classical and quantum information measures can be naturally considered in this matrix approach, which, by the way, accommodates a Lorentz invariant definition of the relevant intrinsic entanglement quantifier, the quantum concurrence \cite{BernardiniEPJP2}. 

In contrast, when a quantizing external field is introduced, the Hamiltonian has an additional dependence on the position variable, and it is not possible to restrict the relevant inner products only to spinorial degrees of freedom. In other words, the momentum is no longer conserved and the matrix model associated to a fixed momentum $\mathbf{k}$ breaks down. To amend this, one turns to a density matrix that takes into account not only discrete, but also continuous degrees of freedom. Fixing the notation for relativistic systems, typical 4-vectors are written as $x^\mu=(t, \mathbf {x}), u^\mu = (\tau, \mathbf {x}), k^\mu = (k_0, \mathbf {k})$, and the $\lambda$-th component of a Dirac bispinor reads
\begin{equation}
	\phi_\lambda (x + u ) = \psi_{\lambda} (\mathbf{x} + \mathbf{u})\exp[-i k_{0} (t + \tau)].
\end{equation}
One consistent framework for describing the qubit system dynamics is given by associating the density matrix of the system with the so-called equal-time Wigner function \cite{PRA2021}, obtained by an energy-average of the covariant Wigner function,\footnote{From the definition of the equal-time Wigner function, one notices that the matrix-valued $\omega$ is not Hermitian, since its components satisfy $\omega_{ \xi \lambda}^\dagger = (\gamma_0)_{\xi \alpha}\,\omega_{ \alpha \beta} \,(\gamma_0)_{\beta \lambda} $. Thus, one simply chooses the charge density $\omega \gamma_0 $, which is Hermitian. } 

\begin{eqnarray}
\omega_{ \xi \lambda} (\mathbf{x},\mathbf{k};t)&=& \int ^{+\infty} _{-\infty} \hspace{-1em} d \mathcal{E} \, W_{\lambda \xi} (x,k) \nonumber \\
&=& \pi^{-1} \sum _{j,m} \exp[ i(k_{0,j} - k_{0,m} ) t] \int d\tau \int ^{+\infty} _{-\infty} \hspace{-1em} d \mathcal{E} \exp[ -i( 2\mathcal{E} - k_{0,j} - k_{0,m} ) \tau] \nonumber \\
&& \quad \times \quad \pi ^{-3} \int d^3 \mathbf{u} \exp[2i \mathbf{k}. \mathbf{u}] \bar{\psi}_{\lambda,j}(\mathbf{x} - \mathbf{u})\psi_{\xi,m}(\mathbf{x} + \mathbf{u}) \nonumber \\
&=& \pi^{-3} \sum _{j,m} \exp[ i(k_{0,j} - k_{0,m}) t] \int d^3 \mathbf{u} \exp[2i \mathbf{k}. \mathbf{u}] \bar{\psi}_{\lambda,j}(\mathbf{x} - \mathbf{u})\psi_{\xi,m}(\mathbf{x} + \mathbf{u}).\label{equaltimewignerfunction}
\end{eqnarray}
It supports a decomposition over phase-space functions that have well-defined transformation rules under Lorentz transformations.\footnote{Such decomposition is possible since there are 16 independent generators of the Clifford Algebra.} The intrinsic information profile is stored in the matrix structure of the Wigner function, where $\bar{\psi} = \psi^* \gamma_0$ and $\psi$ can be regarded in matrix language as row and column vectors, respectively. In this approach, while spinor components correspond to spin-parity degrees of freedom, already present in the free Dirac equation, the phase-space dependence introduces the density matrix for continuous-variable systems. 

\subsection{Linear entropies}

These degrees of freedom are properly handled once the relevant density matrix of each Hilbert space is identified. The spin-parity density matrix is obtained by a phase-space average,
\begin{equation}
 \rho _{SP} = \int\hspace{-.2cm}d^3\mathbf{x}\int\hspace{-.2cm}d^3\mathbf{k} \, \omega(\mathbf{x},\,\mathbf{k};\,t) \gamma^{0},
 \label{density1}
\end{equation}
which is, by the way, also a valid density operator for standard two-qubit systems, since it is Hermitian and can be normalized. However, it is a reduced density matrix where phase-space averaging plays the role of a trace over continuous variables. If the qubit-system is entangled by phase-space variables, then quantum correlations are lost after this integration. Conversely, the phase-space density matrix reads
\begin{equation}
	\rho_{\{\mathbf{x},\mathbf{k}\}} = \Tr[ \omega(\mathbf{x},\mathbf{k})\gamma_0],
	\label{density2}
\end{equation}
where hereafter $\Tr[...]$ is the trace over spinor indices. It is also a reduced density matrix, not exhibiting unitary purity in the general case as well. Both density matrices are needed to evaluate the complete information profile. One requires that the amount of information be evaluated in terms of the linear entropy.

The linearized von Neumann entropy with respect to the spin-parity Hilbert space reads
\begin{equation}\label{entropysp}
	\mathcal{I}^{SP} = 1 - \Tr[\rho_{SP} ^2].
\end{equation}
For continuous-variable systems, the trace operation is replaced by a phase-space integral,
\begin{equation}\label{entropyps}
	\mathcal{I}_{\{\mathbf{x},\mathbf{k}\}} = 1 - (2\pi)^3 \int\hspace{-.2cm}d^3\mathbf{x}\int\hspace{-.2cm}d^3\mathbf{k} \,\rho ^2 (\mathbf{x},\mathbf{k}) ,
\end{equation}
with $\rho (\mathbf{x},\mathbf{k}) = 	\rho_{\{\mathbf{x},\mathbf{k}\}}$, corresponding to the phase-space linear entropy. The normalization factor naturally shows up because the second term on the right-hand side quantifies the quantum purity in the continuous-variable Hilbert space \cite{Case}. Consequently, pure states in the reduced Hilbert space are also zero-entropy states, and no uncertainty or information is stored in the corresponding variables. 

These definitions put on an equal footing the spinor and phase-space degrees of freedom. The linearized entropies are useful for algebraic computations in phase space because the Von Neumann entropy is ill-defined for a localized density matrix. Therefore, the information stored in the equal-time Wigner function can be evaluated in terms of these entropy measures. Adding all of them yields the total mutual information quantifier \cite{BernardiniEPJP},
\begin{equation}\label{mutualinfo}
	M^{SP}_{\{\mathbf{x},\mathbf{k}\}} = \mathcal{I}_{\{\mathbf{x},\mathbf{k}\}} + \mathcal{I}^{SP} - (1 - \mathcal{P}).
\end{equation}
The linear entropy with respect to the global density matrix needs to be subtracted to take into account the loss of information (or entropy) of the density matrix with respect to the total Hilbert space in terms of the quantum purity $\mathcal{P}$, which is now introduced. One notices that the mutual information quantifier is always a non-negative quantity. For mixed states, this measure is degraded due to the uncertainty of the global density matrix and is always less than the sum of the individual entropies with respect to each Hilbert space.

\subsection{Decomposition over pure Wigner functions and quantum purity}

Recalling that the Weyl-Wigner transform in quantum mechanics applied to a pure density matrix $\hat{\rho} = |\psi \rangle \langle \psi |$ yields a quantum phase-space quasidistribution \cite{Case}, 
an equivalent framework can be considered for bispinors, where the equal-time Wigner function is regarded as the density matrix in phase space. Therefore, the one-to-one relationship between the phase and configuration spaces is a mathematical consequence of the Fourier transform. Nevertheless, it is often the case that there is no absolute certainty in the preparation of a pure quantum state. Then, the density matrix picture becomes inevitable. 

One obtains a straightforward extension to statistical ensembles by studying Wigner functions that allow a decomposition over pure solutions,
\begin{equation}
	W (\mathbf{x},\mathbf{k})= \sum_i p_i \, \omega_i (\mathbf{x},\mathbf{k})
	\label{mix0}
\end{equation}
where $\omega_i (\mathbf{x},\mathbf{k})$ is a stationary pure Wigner function and $\sum_i p_i = 1$. One assumes that each $p_i$ is known a priori; however, all statistical ensembles of pure Wigner functions, stationary or not, can be decomposed in this manner. Since the most general Wigner function is both mixed and nonstationary, extracting each $p_i$ with respect to an infinite dimensional basis is not practical. Instead, this assumption simplifies the study of statistical ensembles without the complicated time dependence in the relativistic framework, which has been discussed elsewhere \cite{PRA2023}. As it will be shown in the next section, the probability distribution turns out to be unique for spinors with some kind of a localizing behavior. 

One cannot write the stationary density matrix $ W (\mathbf{x},\mathbf{k})$ in the form of Eq.~\eqref{equaltimewignerfunction}, since no pure spinor realizes this Wigner function. This is in agreement with the prescription of a general density matrix in terms of pure quantum states $\vert \psi_i \rangle$ of a Hilbert space,
\begin{equation}
	\hat{ \rho} = \sum_i p_i \vert \psi_i \rangle \langle \psi_i \vert.
	\label{densitymatrix}
\end{equation}
In either case, $p_i$ is non-negative and can be regarded as a probability distribution. In fact, one retrieves pure state solutions when all, except one, $p_i = 0$. For the equal-time Wigner function considered here, it has been noticed that this is satisfied when the purity quantum operator, 
\small\begin{eqnarray}
	\mathcal{P}&=& (2\pi)^3\int\hspace{-.2cm}d^3\mathbf{x}\int\hspace{-.2cm}d^3\mathbf{k} \, \Tr\left[\left(\gamma^{0}\omega(\mathbf{x},\,\mathbf{k};\,t)\right)^2\right] = 
	(2\pi)^3\int\hspace{-.2cm}d^3\mathbf{x}\int\hspace{-.2cm}d^3\mathbf{k} \, \Tr\left[\omega(\mathbf{x},\,\mathbf{k};\,t)\,\omega^{\dagger}(\mathbf{x},\,\mathbf{k};\,t)\right],
	\label{puritydef}
\end{eqnarray}\normalsize
is maximized to unity \cite{BernardiniEPJP}. In a similar fashion to qubit systems, one calculates the quantum purity by tracing over the relevant degrees of freedom of the density matrix squared. However, for confined systems, the complete trace operator includes the trace over continuous degrees of freedom, i.e. phase-space integration, and thus the pure-state constraint $\mathcal{P}=1$ cannot be verified without averaging over phase-space variables. From now on, one omits the general time-dependence of the quantum purity since only stationary solutions will be discussed.

The orthogonality of stationary solutions is retained in phase space in the sense of Eq.~\eqref{puritydef}, simplifying the quantum purity expression for particular mixed solutions. For any two pure stationary solutions of a given Hamiltonian, 
\begin{equation}
	W(\mathbf{x},\mathbf{k})= p_A \, \omega ^A (\mathbf{x},\mathbf{k}) + p_B \, \omega ^B (\mathbf{x},\mathbf{k})
	\label{mixing}
\end{equation}
is also a stationary solution of the same Hamiltonian (cf. (\ref{equaltimewignerfunction})), and one uses Eq.~\eqref{puritydef} to calculate the purity of the mixed state,
\begin{equation}
	\mathcal{P}[W] = p_A ^2 + p_B ^2 + 2 p_A p_B \chi_{AB},
	\label{puritydem}
\end{equation}
where 
\begin{equation}
	\chi_{AB} = (2\pi)^3\int\hspace{-.2cm}d^3\mathbf{x}\int\hspace{-.2cm}d^3\mathbf{k} \, \Tr\left[\omega_B(\mathbf{x},\,\mathbf{k})\,\omega_A ^{\dagger}(\mathbf{x},\,\mathbf{k})\right]. \nonumber
\end{equation}
Only $\chi_{AB} = 0$ is valid, which follows from the orthogonality of the pure state solutions in configuration space. The quantum purity depends exclusively on the weight parameters $p_A$ and $p_B$, which is straightforwardly generalized to a linear combination of multiple Wigner functions. Thus, two pure states do not overlap in phase space on average, and the quantum purity always satisfies $1/m \leq\mathcal{P} \leq 1$ for a statistical mixture of $m$ pure Wigner functions, where the lower bound is obtained for an equal mixture. 

The point to be made from all the entropy measures is that the formulas for correlations in two-qubit systems could be naively applied to the phase-space averaged density matrix from Eq.~\eqref{density1}, a reduced density matrix. For instance, one considers the concurrence quantifier. Even though separability criteria are verified by standard matrix diagonalization, measures of correlations thus obtained yield, at best, a lower bound on the intrinsic quantum correlation structure of Dirac bispinors. At worst, the eigenvalues that appear in the concurrence formula from Eq.~\eqref{qubitconcurrence} are ambiguous for mixed Wigner functions because the square root of the continuous-variable density matrix is ill-defined. 

In order to obtain an explicit expression for the quantum concurrence in a confining external field, it is argued that the quantization of the Wigner function carries on to the correlation structure as well. In this case, the expression for the quantum concurrence is also quantized. Thus, there is a reduced set of pure states that realize this particular mixed state, and optimization algorithms are rather simplified. This is possible only when the energy spectrum is discrete, and the relevant density matrix is also labeled by additional quantum numbers. In a magnetic field, these quantum numbers correspond to Landau level indices, which immediately restrict the set of pure states in the decomposition of density matrices.

In the next section, quantum and classical correlations of relativistic Landau levels shall be measured. To exhibit the distinguishing features for mixed states, superpositions and mixtures must be contrasted in the framework of Dirac-like Wigner functions. Hence, the correlation structure of mixed Gaussian states can be more properly discussed.

\section{Dirac mixtures in a magnetic field}

Considering the extension from pure to mixed states, a set of stationary Wigner matrices (cf. \eqref{equaltimewignerfunction}) can be yielded from preliminarily reported \cite{BernardiniEPJP} phase-space stationary solutions. The quantization scheme is featured by applying a constant magnetic field, which leads to an infinite set of solutions labeled by the quantum number $n$, in correspondence to the so-called Landau levels. For each $n\neq0$ there are four different solutions; otherwise, there are two solutions corresponding to the ground state \cite{BernardiniEPJP,PRA2021}. The Hamiltonian is given by

\begin{equation}
H_{\rm B}= \mbox{\boldmath$\alpha$} \cdot ({\bf p} + (-1)^r\, e {\bf A}) + \beta m,
\label{eq0}
\end{equation}
where $r=1,2$ corresponds to negatively and positively charged particles, respectively. Here, the charge of the particle can also be used to label the intrinsic parity, a qubit state, of the particle. The gauge is chosen as ${\bf A} = \mathcal{B}\,x\, \hat{\bf y}$, since the nontrivial continuous degrees of freedom are reduced to one dimension. Nevertheless, any relevant result should not depend on this particular gauge. Only the magnetic field ${\bf B} = \mbox{\boldmath$\nabla$} \times {\bf A}= \mathcal{B} \, \hat{\bf z}$ along the $z$-direction and physical parameters should influence observables.

The phase-space eigensolutions describing the mixed solutions are described in Appendix \ref{AppA}. In particular, one notices that no relevant information is stored in the plane wave contribution, which is always integrated out of all expressions. The set of solutions at each $n$ Landau level corresponds to matrix-valued Wigner functions that carry additional labels with respect to the intrinsic parity, $r=1,2$, and the spin projection (helicity), $\pm$. 

In order to study superpositions and mixtures in phase space, it is essential to fix the center of the motion $s_{r=1}\equiv s$. In this case, the set for $E > 0$ is mapped to positive parity states, $r=1$, with
\begin{eqnarray}
	\omega^+_{n,1}(s,\,k_x) &=&\eta_{n}\left( \begin{array}{cccc} 
		\mathcal{L}_{n-1} & 0 & -A_{n}\,\mathcal{L}_{n-1} & B_{n}\,\mathcal{M}_{n} \\ 
		0 & 0& 0& 0\\
		A_{n}\,\mathcal{L}_{n-1} & 0 & -A^2_{n}\,\mathcal{L}_{n-1} & A_{n}B_{n}\,\mathcal{M}_{n} \\ 
		-B_{n}\,\mathcal{M}_{n} & 0 & A_{n}B_{n}\,\mathcal{M}_{n}& -B^2_{n}\,\mathcal{L}_{n} 
	\end{array} \right), 
\label{9996}
\end{eqnarray}
\begin{eqnarray}
	\omega^-_{n,1}(s,\,k_x) &=& \eta_{n}\left( \begin{array}{cccc} 
		0 & 0& 0& 0\\
		0 & \mathcal{L}_{n} & B_{n}\,\mathcal{M}_{n}& A_{n}\,\mathcal{L}_{n} \\ 
		0 & -B_{n}\,\mathcal{M}_{n} & -B^2_{n}\,\mathcal{L}_{n-1}& -A_{n}B_{n}\,\mathcal{M}_{n} \\ 
		0 & -A_{n}\,\mathcal{L}_{n} & -A_{n}B_{n}\,\mathcal{M}_{n}& -A^2_{n}\,\mathcal{L}_{n} \\ 
	\end{array} \right),
\label{9997}
\end{eqnarray}
whereas the set for $E<0$ is mapped to negative parity states, $r=2$, with
\begin{eqnarray}
	\omega^-_{n,2}(s,\,k_x) &=& \eta_{n}\left( \begin{array}{cccc} 
		A^2_{n} \mathcal{L} _{n-1} & A_{n}B_{n}\,\mathcal{M} _{n} & -A_{n}\,\mathcal{L} _{n-1} & 0 \\ 
		A_{n}B_{n}\,\mathcal{M} _{n} & B^2_{n}\,\mathcal{L} _{n} & -B_{n}\,\mathcal{M} _{n} & 0 \\ 
		A_{n}\,\mathcal{L} _{n-1} & B_{n}\,\mathcal{M} _{n} & -\mathcal{L} _{n-1} & 0 \\ 
		0 & 0& 0& 0
	\end{array} \right).
\label{9998}
\end{eqnarray}

\begin{eqnarray}
	\omega^+_{n,2}(s,\,k_x) &=& \eta_{n}\left( \begin{array}{cccc} 
		B^2_{n} \mathcal{L} _{n-1} & - A_{n}B_{n}\,\mathcal{M} _{n} & 0 & -B_{n}\,\mathcal{M} _{n} \\ 
		- A_{n}B_{n}\,\mathcal{M} _{n} & A^2_{n}\,\mathcal{L} _{n} & 0 &A_{n}\,\mathcal{L} _{n}\\ 
		0 & 0& 0& 0\\
		B_{n}\,\mathcal{M} _{n} & -A_{n}\,\mathcal{L} _{n} &0 & -\mathcal{L} _{n} \\ 
	\end{array} \right),
	\label{9999}
\end{eqnarray}
where $A_n, B_n, \eta_n$ are non-negative constant parameters (smaller than unity) related to the one-particle parameters,
\begin{equation}\label{parameters}
	A_{n}\, = \frac{k_z}{E_{n} +m},\,\,B_{n}\, = \frac{\sqrt{2n\,e\mathcal{B}}}{E_{n} +m},\,\,\,\mbox{and} \,\,\, \eta_{n}=\frac{E_{n} +m}{2E_{n}}
\end{equation}
given in terms of the energy eigenvalues $E_n = \sqrt{m^2 + k_z^2 + 2n\,e\mathcal{B}}$. The correspondence with the harmonic oscillator (HO) basis is identified by the phase-space eigenfunctions, 
\begin{equation}\label{norr01}
	\mathcal{L}_{n} \equiv \mathcal{L}_{n}(s,\,k_x) = \frac{(-1)^{n}\sqrt{e\mathcal{B}}}{\pi} \,\exp\left[-(s^2 + k^2_x)\right]\, {L}_{n}\left[2(s^2 + k^2_x)\right],
\end{equation}
with $L_{n}$ corresponding to a Laguerre polynomial, such that
\begin{eqnarray}
	\int_{_{-\infty}}^{^{+\infty}}\hspace{-.5 cm}dx\,\int_{_{-\infty}}^{^{+\infty}}\hspace{-.5 cm}dk_x \,\mathcal{L}_{n}(s,\,k_x) &=& 1 \label{norr0w} \\ \frac{2\pi}{\sqrt{e\mathcal{B}}}\int_{_{-\infty}}^{^{+\infty}}\hspace{-.5 cm}dx\,\int_{_{-\infty}}^{^{+\infty}}\hspace{-.5 cm}dk_x \,\mathcal{L}_{n}(s,\,k_x)\,\mathcal{L}_{m}(s,\,k_x) &=& \delta_{mn}\label{norr0w1},
\end{eqnarray}
with $ds = \sqrt{e\mathcal{B}}\, dx$ and
\begin{eqnarray}
	\mathcal{M}^{(r)}_{n} &\equiv& \mathcal{M}_{n}(s,\,k_x)\nonumber\\
	&=& \frac{(-1)^{n}}{2\pi} \sqrt{\frac{ e\mathcal{B}}{ n}}\,\exp\left[-(s^2 + k^2_x)\right] \left(\frac{d}{ds}{L}_{n}\left[2(s^2 + k^2_x)\right]\right) \label{norr02},
\end{eqnarray}
which averages out to zero,
\begin{equation}
	\int_{_{-\infty}}^{^{+\infty}}\hspace{-.5 cm}dx\,\int_{_{-\infty}}^{^{+\infty}}\hspace{-.5 cm}dk_x \,\mathcal{M}_{n}(s,\,k_x) = 0,\label{norr0w2}
\end{equation}
and is also orthogonal to all $\mathcal{L}_{n}(s,\,k_x)$ (cf. Eq.~\eqref{norr0w1}). The HO basis is a complete basis in phase space, and thus phase-space integrals are straightforwardly calculated using such orthogonality relations.

\subsection{Admissible Wigner functions}
Two additional properties of Dirac-like Wigner functions must be addressed. They follow from the Weyl transform applied to the relevant density matrices and they will be relevant for evaluating entropy measures.
 
All Wigner functions are normalized to unity since the probability is conserved. The normalization condition applied to Dirac bispinors encompasses both the trace over discrete degrees of freedom, $\Tr[...]$, and the trace over continuous degrees of freedom, i.e.
\begin{equation}
	\int^{+\infty}_{-\infty} \hspace{-.5cm} {dx}\int^{+\infty}_{-\infty} \hspace{-.5cm}{dk_x}\,\Tr [(\omega^s _{n,r} (s,\, k_x) \gamma_0)] = 1,
	\label{normalization}
\end{equation}
of course, for both pure and mixed states.

The pure-state constraint is satisfied by eigenstates of a given Hamiltonian. Indeed, the basis introduced above is composed by pure states,

\begin{equation}
	\frac{2\pi}{\sqrt{e\mathcal{B}}} \,	\int^{+\infty}_{-\infty} \hspace{-.5cm} {dx}\int^{+\infty}_{-\infty} \hspace{-.5cm}{dk_x}\,\Tr [(\omega^s _{n,r} (s,\, k_x) \gamma_0)(\omega^{s'} _{n',r'} (s,\, k_x) \gamma_0)] = \delta_{s,s'} \delta_{r,r'}\delta_{n,n'},
	\label{orthogonality}
\end{equation}
and the orthogonality property is extended to Dirac-like Wigner functions. The additional factor comes from the normalization of the phase-space functions (cf. (\ref{norr0w1})). This confirms that orthogonal wave functions remain orthogonal in phase space. In addition, Eq.~\eqref{orthogonality} suggests that particular mixed states can be expanded in phase space with positive coefficients (cf. (\ref{mixing})). However, this is not possible for a general $4 \times 4$ density matrix. Namely, a superposition needs an additional term that must be calculated in configuration space. To verify such an assertion, an instructive example shall be discussed in the following, distinguishing typical quantum phenomena from classical ones due to the mixture of states.

\subsection{Randomness and interference effects in quantum correlations}
To depict the differences between superpositions and mixtures within the Weyl-Wigner framework, it is useful to consider first a quantum superposition within the degenerate subspace, 
\begin{eqnarray} \label{superposition}
 \omega_{n,\theta} &=& \sin^2 (\theta) \, \omega ^+ _{n,1} + \cos^2 (\theta) \, \omega ^- _{n,1} + \sin(\theta) \cos(\theta) \, \Omega_n, \\
 \mbox{where} \quad \Omega_n (s,\,k_x) &=& \eta_{n}\left( \begin{array}{cccc} 
0& \mathcal{M}_{n} & -B_{n}\,\mathcal{L}_{n-1} & A_{n}\,\mathcal{M}_{n} \\ 
\mathcal{M}_{n} & 0& -A_{n}\,\mathcal{M}_{n}& -B_{n}\,\mathcal{L}_{n} \\
B_{n}\,\mathcal{L}_{n-1} & A_{n}\mathcal{M}_{n} & 2A_{n}B_{n}\,\mathcal{L}_{n-1} & (A_{n}^2 - B_{n} ^2)\,\mathcal{M}_{n} \\ 
-A_{n}\,\mathcal{M}_{n} & -B_{n}\,\mathcal{L}_{n} & (A_{n}^2 - B_{n} ^2)\,\mathcal{M}_{n}& -2A_{n}B_{n}\,\mathcal{L}_{n} 
\end{array} \right) \nonumber,
\end{eqnarray}
which is obtained by plugging the normalized pure states $\sin (\theta) \, u^+ _{n,1}(s) + \cos (\theta) \, u^- _{n,1}(s)$ (cf. Appendix \ref{AppA}) into the definition of the equal-time Wigner function in Eq.~\eqref{equaltimewignerfunction}. Thus, the first two terms on the right-hand side stand for the density matrix of each state in the superposition, and the third term corresponds to an overlap between them, where $\sin(\theta) \cos(\theta)$ naturally quantifies this overlap. Indeed, $ \Omega_n (s,\,k_x)$ is by no means a valid Wigner function, since it cannot be normalized (cf. (\ref{normalization})),
\begin{equation}
\int^{+\infty}_{-\infty} \hspace{-.5cm} {dx}\int^{+\infty}_{-\infty}\hspace{-.5cm}{dk_x}\,\Tr[\Omega_n (s,\,k_x)\gamma_0] = -2\eta_{n} \, A_{n}B_{n} \int^{+\infty}_{-\infty} \hspace{-.5cm} {dx}\int^{+\infty}_{-\infty}\hspace{-.5cm}{dk_x}\,(\mathcal{L}_{n} - \mathcal{L}_{n-1}) = 0,
\end{equation}
and $\sin(2\theta)/2$ is negative for $\pi/2<\theta < \pi$; nevertheless, the complete expression in \eqref{superposition} is in fact a valid Wigner function. First, it is normalized,
\begin{eqnarray}
	\int^{+\infty}_{-\infty} \hspace{-.5cm} {dx}\int^{+\infty}_{-\infty}\hspace{-.5cm}{dk_x}\,\Tr[\omega_{n,\theta} (s,\,k_x)\gamma_0] &=& \int^{+\infty}_{-\infty} \hspace{-.5cm} {dx}\int^{+\infty}_{-\infty}\hspace{-.5cm}{dk_x}\,\Tr\bigg[ \bigg( \sin^2 (\theta) \, \omega ^+ _{n,1} + \cos^2 (\theta) \, \omega ^- _{n,1} \bigg) \gamma_0\bigg] \nonumber \\
	&=& 1,
\end{eqnarray}
where one has used the normalization condition in Eq.~\eqref{normalization} for each term above. Second, the matrix $\Omega_n (s,\,k_x)$ drives the interference effects between the pure state solutions $\omega ^+ _{n,1}$ and $\omega ^- _{n,1} $, not affecting the purity content of the superposition, which is evaluated as 
 \begin{eqnarray}
\mathcal{P} &=& \frac{2\pi}{\sqrt{e\mathcal{B}}}\int^{+\infty}_{-\infty} \hspace{-.5cm} {dx}\int^{+\infty}_{-\infty} \hspace{-.5cm}{dk_x}\,\Tr [(\omega_{n,\theta} (s,\, k_x) \gamma_0)^2]
\nonumber \\
&=& (\sin^2 (\theta) + \cos^2 (\theta))^2 \eta_n ^2 ( 1 + A_n ^2 + B_n ^2)^2 \nonumber \\
&=& 1,
\end{eqnarray}
i.e. as a pure state. The simplification in the last line, $\eta_n ( 1 + A_n ^2 + B_n ^2) = 1$ (cf. \eqref{parameters}) follows from an algebraic manipulation of the energy parameters. Thus, Eq.~\eqref{superposition} is not a mixture of quantum states. It corresponds to a change of basis since the spectrum has a two-fold spin degeneracy. 

On the other hand, a true mixture between two stationary Wigner functions of a fixed quantum number $n$ reads
\begin{equation}
 \omega_{n,\phi}= \sin^2(\phi) \, \omega ^+ _{n,1} + \cos^2(\phi) \, \omega ^- _{n,1}.
 \label{mixture1}
\end{equation}
It is noticed that the ensemble is also constituted by states within the degenerate subspace, and only positive coefficients appear in the decomposition, without any interference term. The purity content is similarly evaluated,
 \begin{eqnarray}
\mathcal{P} &=& \frac{2\pi}{\sqrt{e\mathcal{B}}} \int^{+\infty}_{-\infty} \hspace{-.5cm} {dx}\int^{+\infty}_{-\infty} \hspace{-.5cm}{dk_x}\,\Tr [(\omega_{n,\phi} (s,\, k_x) \gamma_0)^2]
 \nonumber \\
&=& (\sin^4 (\phi )+ \cos^4 (\phi))\eta_n ^2 ( 1 + A_n ^2 + B_n ^2)^2 \nonumber \\
&=& \sin^4 (\phi )+ \cos^4 (\phi) \leq 1.
\end{eqnarray}
Thus, $\phi$ is a true mixing parameter and yields pure state solutions for $\phi = l \pi/2$ with integer $l$. One notices that the purity only depends on the probability distribution itself, as expected from Eq.~\eqref{puritydem}. In brief, the quantum purity measure exclusively quantifies the effects of linear combinations that take place in phase space but is not affected by superpositions in configuration space. 

In regards to the probability distribution uniqueness, i.e. if it is possible to obtain the same mixed state (cf. (\ref{mixture1})) with a distinct decomposition, one first notices that the HO basis is orthogonal in phase space. Thus, only mixtures for a fixed $n$ from Eqs.~(\ref{9996})-(\ref{9999}) exhibit the appropriate scalar functions. Even then, one now has to match all the 16 constants from the components of the density matrix. It is straightforward to verify that it is not possible to obtain the same state with a distinct linear combination, with positive coefficients, of Wigner functions with other pure states. Therefore, this decomposition is unique. 

\begin{figure} [H]
	\centering 
	\includegraphics[width=0.9\textwidth]{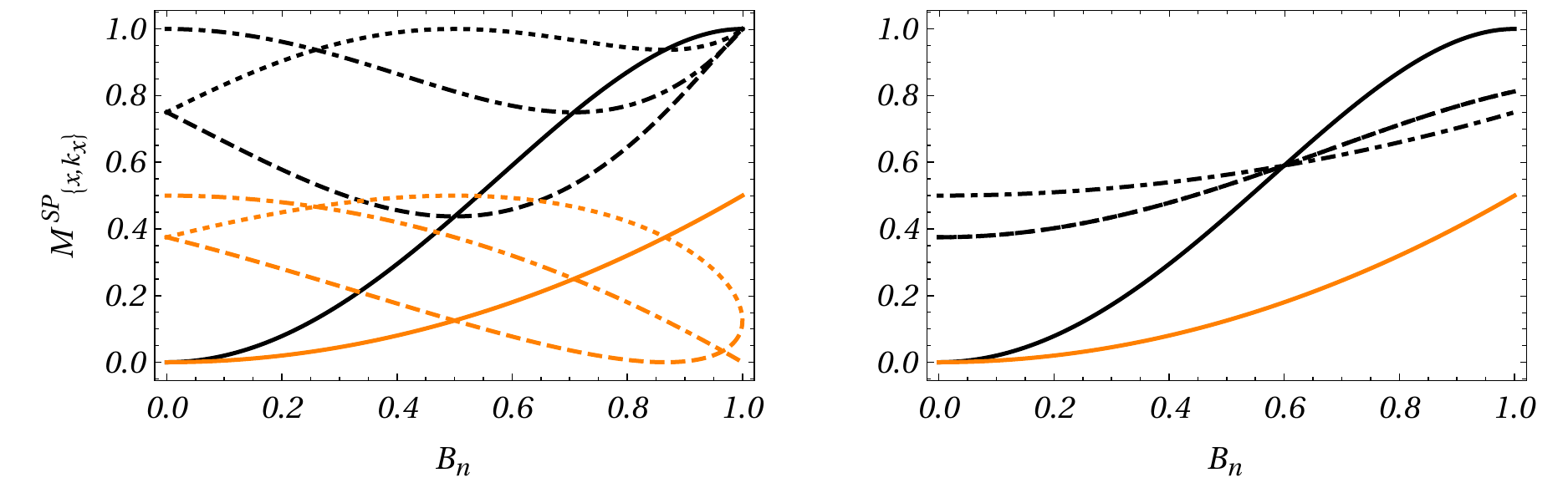} 
	\caption{(Color online) Total intrinsic information for stationary Landau levels. The amount of information is computed through the mutual information quantifier (cf. Eq.~\eqref{mutualinfo}; black lines) for which it is assumed that $m=0$ (cf. Eq.~\eqref{parameters}). (Left plot) Results are for pure superpositions of spin up and down states from Eq.~\eqref{superposition}, for coefficients $\sin^2(\theta) = 0$ (solid), $1/4$ (dashed), $1/2$ (dot-dashed), and $3/4$ (dotted). (Right plot) Results are for mixtures of spin up and down states from \eqref{mixture1}, for coefficients $\sin^2 (\phi) = 0$ (solid), $1/4$ (dashed), $1/2$ (dot-dashed), and $3/4$ (dotted), i.e. with the same line patterns. 
Results for the intrinsic concurrence (orange lines) are also depicted in both plots, with the same line patterns for the related coefficients.	
Whereas some amount of information is lost in mixed states for strong magnetic fields ($B_n =1$), randomness creates correlations in weak magnetic fields ($B_n = 0$). In addition, all mixtures share the same value of quantum concurrence.}
	\label{fig1}
\end{figure}

One now evaluates the mutual information expression in \eqref{mutualinfo} for superposition and mixing angles $\theta$ and $\phi$, respectively. Although the cumbersome dependence on parameters makes this measure somewhat obscure to interpret, one observes more illuminating features if results are instead depicted in Fig.~\eqref{fig1} for coincident values of $\theta$ and $\phi$. The same angles were chosen so as to compare each mixture with a corresponding quantum superposition, such that in both cases one obtains a stationary Wigner function since energy levels are degenerate. 
The external field is a factor that strongly affects correlations. On the one hand, the mutual information for quantum superpositions $(\theta \neq 0)$ is a non-monotonic function of the magnetic field represented by $B_n = \frac{\sqrt{2n\,e\mathcal{B}}}{\sqrt{ k_z^2 + 2n\,e\mathcal{B}}}$ because the interference pattern could either amplify or suppress its effects on correlations. It is somewhat surprising, though, that for a strong field ($B_n \approx 1$), this quantifier approaches unity for all values of $\theta$. On the other hand, correlations for a fixed $\phi$ always increase when the magnetic field increases. One notices, however, that for a fixed value of the interaction, $\phi$ tunes the total amount of information.

In fact, there is an optimal value of the field that determines whether increasing $\phi$ degrades or increases these correlations. In the figure, this corresponds to $B_n \approx 0.6$. Below this value, mixing creates correlations; above this value, mixing suppresses correlations. For instance, considering the pure state as the reference state, an equal mixture with $\sin^2 (\phi) = 1/2$ (dot-dashed lines) loses almost one quarter of correlations due to randomness in a strong magnetic field. However, in a weak magnetic field ($B_n = 0$), half of these correlations are created for the same mixing angle.

Additional features are observed as well. Two distinct mixtures for which the values of $\phi$ are complementary exhibit coincident values for all quantifiers. In particular, for $\sin^2 (\phi)= 1/4 \,(\phi = \pi/6), \,3/4 \, (\phi = \pi/3)$, one obtains the same information measures. This means that swapping the probability distribution of a mixture of two states generates an equivalent ensemble. In contrast, there is no equivalence between quantum superpositions with the same angles $\sin^2 (\theta)= 1/4, 3/4$ (dashed and dotted lines, respectively). This is a consequence of the fundamental distinction between superpositions and mixtures. Whereas $\theta$ quantifies the interference of states, the mixing angle $\phi$ only attributes a probability, or chance, of measuring each pure state. 

\subsubsection{Spin-parity concurrence in Landau levels}

Aside from the total mutual information measured by the associated linear entropies, quantum concurrence can also be computed. However, the exact formula for two-qubit systems correlated by phase-space variables has been demonstrated only for pure states, as presented in the Appendix \ref{AppB}. The main result is that the additional degree of freedom related to the orbital wave function introduces a quantization in the pure-state entanglement, which exhibits coincident values for any of the density matrices shown in Eqs.~(\ref{9996})-(\ref{9999}),

\begin{equation}\label{concurrence}
	\int_{_{-\infty}}^{^{+\infty}}\hspace{-.5 cm}ds\,\int_{_{-\infty}}^{^{+\infty}}\hspace{-.5 cm}dk_x \ \mathcal{C}^2 [\omega^{\pm}_{n,r}] = 2 (\eta_n B_n ) ^2 ,
\end{equation}
where these coefficients were introduced in Eq.~\eqref{parameters} and are related to the ratio between the magnetic field intensity and the one-particle energy eigenvalue. This means that concurrence is quantized, and all quantum states of the same Landau level exhibit coincident values of the spin-parity entanglement. 

Returning to the statistical ensemble parameterized by the angle $\phi$ in Eq.~\eqref{mixture1}, one now recalls that the quantum concurrence is obtained by looking for the minimal value among all sets that realize this state (cf. \eqref{averagedconcurrence}). It is fortunate, however, that these mixed states are in fact unique, as discussed above. Thus,
 \begin{eqnarray}
 	\int_{_{-\infty}}^{^{+\infty}}\hspace{-.5 cm}ds\,\int_{_{-\infty}}^{^{+\infty}}\hspace{-.5 cm}dk_x \ \mathcal{C}^2 [\omega_{n,\phi}] &=& (\sin^2 (\phi) + \cos^2 (\phi)) 	\int_{_{-\infty}}^{^{+\infty}}\hspace{-.5 cm}ds\,\int_{_{-\infty}}^{^{+\infty}}\hspace{-.5 cm}dk_x \ \mathcal{C}^2 [\omega^\pm _{n}] \nonumber \\
 	&=& 2 (\eta_n B_n ) ^2,
 \end{eqnarray}
independent of the mixing angle.

On the other hand, the superposition considered in Eq.~\eqref{superposition} has a concurrence that can be written as (cf. Appendix \ref{AppB}) 

\begin{equation}
\int_{_{-\infty}}^{^{+\infty}}\hspace{-.5 cm}ds\,\int_{_{-\infty}}^{^{+\infty}}\hspace{-.5 cm}dk_x \ \mathcal{C}^2 [\omega_{n,\theta}] = 2\eta_n ^2 \, \big(B_n \cos(2\theta) - A_n \sin(2\theta)\big)^2,
\end{equation}
simplifying into the quantum concurrence in Eq.~\eqref{concurrence} when $\cos(2\theta)=\pm 1$. Hence, this quantifier distinguishes between a quantum superposition and a statistical ensemble. 

This is also included in Fig.~\eqref{fig1} for $A_n ^2 + B_n ^2 =1$ (massless limit). The results indicate the most dramatic difference between pure and mixed states in a quantizing field. While a complete disentanglement is observed for quantum superpositions with $\sin^2(\theta) = 1/4, 1/2$ (dashed and dot-dashed lines, respectively), the intrinsic entanglement profile is unaffected by mixing states that have the same intrinsic concurrence. In fact, this measure has the same qualitative behavior as the mutual information for mixed states, increasing whenever the magnetic field increases. Therefore, quantum concurrence is not sensitive to the uncertainty associated to the quantum state preparation.

Summarizing, it has been evinced that the total information stored in discrete and continuous degrees of freedom, quantified by the mutual information, is affected by randomness and interference in different ways. Whereas the former limits the maximum information that can be measured in the mixed state, the latter changes the (monotonic) dependence on the magnetic field for a superposition. In regards to quantum correlations, they are determined by interference effects, not by a phase-space probability distribution. This implies that the mutual information minus the quantum concurrence quantifies a type of correlation that can be augmented or reduced by randomness. In contrast, interference affects all types of correlations in a nontrivial manner. 

In the next subsection, one evaluates correlations of mixed Gaussian states, which have similarities with classical distributions in phase space. Namely, the probability distribution is always positive. They are obtained by inspecting the generating functions of Laguerre polynomials in phase space.

\subsection{Gaussian mixed states}
\subsubsection{Maximally mixed Landau level}
One first recalls that there are four pure states for each principal quantum number. Thus, the maximally mixed state within this subspace is obtained by an equal mixture with both signs of the parity,
\begin{eqnarray}
	W_n(s,\,k_x) &=& \frac{1}{4} \sum_{s,r=1} ^2 \omega^s _{n,r} (s,\,k_x), \nonumber \\
	&=& \frac{1}{4} \left( \begin{array}{cccc} 
		\mathcal{L}_{n-1} & 0& 0& 0\\
		0 & \mathcal{L}_{n} &0 & 0\\ 
		0 & 0 &\,-\mathcal{L}_{n-1}&0 \\ 
		0 & 0& 0& -\mathcal{L}_{n} \\ 
	\end{array} \right),
	\label{mix1}
\end{eqnarray}
where the restricted sum over $s=1$ yields the results from the previous section. However, those states were difficult to manipulate because the density matrix is not diagonal and depends on several physical parameters. Here, however, the rather simple form of the Wigner function simplifies all quantifiers. 

The purity is straightforwardly computed since the density matrix is diagonal,
\begin{eqnarray}
\mathcal{P} &=& \frac{2\pi}{\sqrt{e\mathcal{B}}} \int^{+\infty}_{-\infty} \hspace{-.5cm} {dx}\int^{+\infty}_{-\infty} \hspace{-.5cm}{dk_x}\,\Tr [(W_n (x,\, k_x) \gamma_0)^2] \nonumber \\
&=& \frac{\pi}{4\sqrt{e\mathcal{B}}}\int^{+\infty}_{-\infty} \hspace{-.5cm} {dx}\int^{+\infty}_{-\infty} \hspace{-.5cm}{dk_x} \left( \mathcal{L}_{n-1} ^2 + \mathcal{L}_{n} ^2 \right) \nonumber \\
&=& \frac{1}{4},
\label{purity}
\end{eqnarray}
where one has obtained the last equality by using Eq.~\eqref{norr0w1}. Thus, this mixed state indeed corresponds to an equal mixture of four pure states.

Moving to the computation of entropies with respect to phase-space and spin-parity space, one has
\begin{eqnarray}
\mathcal{I}_{\{x,k_x\}} &=& 1 - \frac{2\pi}{\sqrt{e\mathcal{B}}}\int_{_{-\infty}}^{^{+\infty}}\hspace{-.5 cm}dx\,\int_{_{-\infty}}^{^{+\infty}}\hspace{-.5 cm}dk_x 
\Tr\left[\gamma_0 W_n (s(x),\,k_x)\right]^2=
\frac{1}{2},
\end{eqnarray}
and 
\begin{eqnarray}
\mathcal{I}^{SP} &=& 1 - \Tr\left[\left(\gamma_{0}\,\int_{_{-\infty}}^{^{+\infty}}\hspace{-.5 cm}dx\,\int_{_{-\infty}}^{^{+\infty}}\hspace{-.5 cm}dk_x \,W_n (s,\,k_x)\right)^2\right]=
\frac{3}{4},
\end{eqnarray}
respectively. These are calculated by considering the orthogonality relations of the basis functions in Eqs.~(\ref{norr0w})-(\ref{norr0w1}). Then, the total mutual information reads
\begin{eqnarray}
M^{SP}_{\{x,k_x\}} &=& \mathcal{I}_{\{x,k_x\}} + \mathcal{I}^{SP} - (1 - \mathcal{P})\nonumber\\
&=& \frac{1}{2}.
\end{eqnarray}
Surprisingly, it does not depend on any particular regime or energy parameter, and all correlations are constant for the maximal mixture within the same Landau level. This can be compared with the intrinsic (squared) concurrence previously discussed, 
\begin{equation}
 \mathcal{C}^2 = 2 (\eta_n B_n ) ^2 ,
 \label{concurrence1a}
\end{equation}
which is a quantity that takes a fixed value for each Landau level and is constrained to $ 0 \leq \mathcal{C}^2 \leq 1/2$ (cf. \eqref{parameters}). Therefore, at the lower bound, the total information is dominated by classical correlations. At the upper bound, the amount of information is dominated by concurrence, and randomness degrades all classical correlations. 
\subsubsection{Mixing distinct Landau levels}
Using this diagonal state for each Landau level, one now seeks states that are arbitrarily mixed. Generally, they correspond to ensembles of all eigenstates. One assumes that they take the form
\begin{equation}
 \mathcal{W}(s,\,k_x) = \sum_{n=0}^{\infty} p_n \, W_n(s,\,k_x),
\end{equation}
where $W_n(s,\,k_x)$ is an equal mixture of states as obtained in Eq.~\eqref{mix1}.

For appropriate choices of $p_n$, subtle analytic functions can be obtained. Assuming further that $p_n = z^n$ yields a straightforward interpretation for the mixture. In this case, $z=0$ selects only the ground state, whereas $z=1$ selects infinitely many Landau levels. This parameterization is useful because Laguerre polynomials satisfy
\begin{equation}
 \sum_{n=0}^{\infty} L_n (r^2) \, z^n = \frac{1}{1-z} \exp\left(\frac{z\,r^2}{z-1}\right),
\end{equation}
where the right-hand side is their generating function with $r^2 = s^2 + k_x ^2$. Recalling the complete basis functions from \eqref{norr01}, only two components of the Wigner function need to be calculated, 

\begin{eqnarray}
	 \sum_{n=0}^{\infty} 	\mathcal{L}_{n} (s,\,k_x)\, z^n &=& \sqrt{e\mathcal{B}}\,\exp\left(-r^2 \right) \sum_{n=0}^{\infty} (-1)^{n}\, {L}_{n}(2r^2)\,z^n \nonumber \\
	 &=& \sqrt{e\mathcal{B}}\, \frac{\exp\left[2r^2\left(\frac{z}{z+1} -1/2\right)\right]}{z+1},
\end{eqnarray}
and 
\begin{eqnarray}\label{eqaa1}
	\sum_{n=0}^{\infty} 	\mathcal{L}_{n} (s,\,k_x)\, z^{n+1} &=& z \sqrt{e\mathcal{B}}\,\exp\left(-r^2 \right) \sum_{n=0}^{\infty} (-1)^{n}\, {L}_{n}(2r^2)\,z^n \nonumber \\
	&=& z \sqrt{e\mathcal{B}}\, \frac{\exp\left[2r^2\left(\frac{z}{z+1} -1/2\right)\right]}{z+1}.
\end{eqnarray}
Therefore, all components are Gaussian functions in phase space. For admissible Wigner functions and a clear connection with statistical ensembles, $z$ should be restricted to positive values. In fact, $z$ can be regarded as a mixing parameter, selecting states with large quantum numbers as $z \rightarrow 1$. By collecting the normalization factors, the final result is the following Wigner function,
\begin{eqnarray}
 \mathcal{W}_{11} &=& \frac{N}{4}\frac{ \sqrt{e\!B} }{\pi} \frac{z}{z+1}\exp\left[2(s^2 + k_x ^2)\left( \frac{z}{z+1} - 1/2 \right)\right]\label{mixedGaussian1}, \\
 \mathcal{W}_{22} &=&\frac{N}{4} \frac{ \sqrt{e\!B} }{\pi} \left\{\exp[-(s^2 + k_x ^2)] + \frac{1}{z+1}\exp\left[2(s^2 + k_x ^2)\left( \frac{z}{z+1} - 1/2 \right)\right] \right\}\label{mixedGaussian2}, \\
 \mathcal{W}_{33} &=& - \mathcal{W}_{11}, \\
 \mathcal{W}_{44} &=& - \mathcal{W}_{22}, 
\end{eqnarray}
where $\mathcal{W}_{ij}$ denotes the element at the $i$-th row and $j$-th column. Also, $N= (\sum_{n=0}^{\infty} z^n)^{-1}= 1-z $ is the normalization constant, and all other matrix elements are zero\footnote{Before computing the relevant physical observables, it is useful to emphasize that all phase-space integrals are immediately evaluated using polar coordinates since these functions depend exclusively on the phase-space radius $r^2 = s^2 + k_x ^2$, and integrands simplify to exponential functions of a single variable. Noticing that $ds \, dk_x = r \, dr\, d\theta$, 
\begin{eqnarray} \label{gaussianintegration}
	\int_{_{-\infty}}^{^{+\infty}}\hspace{-.5 cm}ds\,\int_{_{-\infty}}^{^{+\infty}}\hspace{-.5 cm}dk_x \, \exp\left[(s^2 + k_x ^2) \alpha \right] &=& 2\pi \,\int_{_{0}}^{^{+\infty}}\hspace{-.5 cm}dr \, r \exp(\alpha\, r^2) \nonumber \\
	&=& -\pi \left(\frac{1}{\alpha} \right),
\end{eqnarray}
for $\Re[\alpha]<0$. All physical observables can be obtained using this change of variables. For instance, the probability density in phase space $\Tr[\mathcal{W} (s,k_x) \gamma_0]$ is a normalized Gaussian function since
{\footnotesize
\begin{eqnarray}
\int_{_{-\infty}}^{^{+\infty}}\hspace{-.5 cm}dx\,\int_{_{-\infty}}^{^{+\infty}}\hspace{-.5 cm}dk_x \,\Tr[\mathcal{W} \gamma_0] &=& (1-z) \int_{_{0}}^{^{+\infty}}\hspace{-.5 cm}dr \, r \left\{ \,\exp\left[2r^2\left( \frac{z}{z+1} - \frac{1}{2}\right)\right] + \exp[-r^2] \right\} \nonumber \\
&=& \label{normalization2} 1.
\end{eqnarray}}}.
\subsubsection{Gaussian quantum purity}
One interesting feature is that as $z$ approaches unity, the probability density spreads over all phase-space coordinates, and the state is no longer localized, which is depicted in Fig.~\eqref{gaussianphasespace}. Since $z=1$ selects all excited levels, the probability becomes evenly distributed over all eigenstates. This can be straightforwardly measured by the quantum purity, which takes a workable form (cf. Appendix \ref{AppC}),
{\footnotesize
\begin{eqnarray}
\mathcal{P}_\mathcal{W} &=& 2\pi\int^{+\infty}_{-\infty} \hspace{-.5cm} {dx}\int^{+\infty}_{-\infty} \hspace{-.5cm}{dk_x}\, \Tr[\left( \gamma_0\mathcal{W} (x,\, k_x) \right) ^2] \nonumber \\ 
&=& \frac{(z-1)(z^2 -2)}{4(z+1)}, \label{puritymixed}
\end{eqnarray}}
where polar coordinates have been used since the exponents are always nonpositive. A vanishing quantum purity for $z=1$ indeed corresponds to a maximally mixed state in the infinite-dimensional Hilbert space. From the probability density in \eqref{normalization2} and the quantum purity above, it could be naively argued that this state is a trivial one. However, the integrated probability still satisfies the normalization condition for any $z$.

\subsubsection{Phase-space uncertainty relations}

\begin{figure} [H]
	\centering
	\includegraphics[width=.8\textwidth]{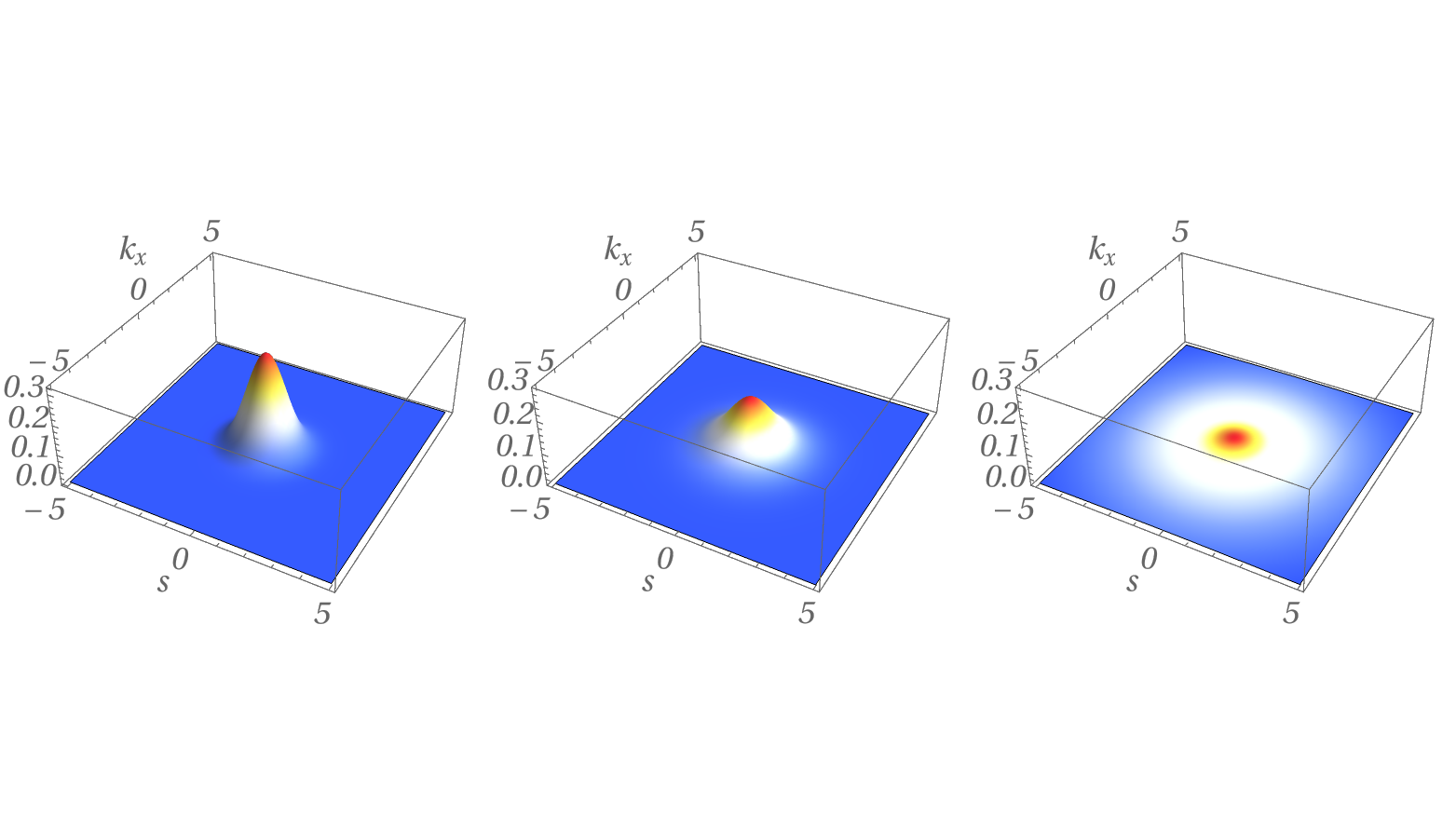}
	\caption{(Color online) Delocalization of mixed Gaussian states in phase space. Localization is measured in terms of the probability density $\frac{\Tr[\mathcal{W} \gamma_0]}{\sqrt{e\mathcal{B}}}$ of states obtained in Eqs.~(\ref{mixedGaussian1})-(\ref{mixedGaussian2}).The so-called coherent state (left plot), with $z=0$, covers the minimum area, whereas the Gaussian states with $z=0.5$ (center plot) and $z=0.9$ (right plot) are smoothed over phase space. }\label{gaussianphasespace}
\end{figure}

One notices that localization in phase space, indirectly controlled by the mixing parameter, cannot be set to arbitrary values. This property is bound to satisfy the uncertainty principle of canonically conjugate variables. Using the (nonintegrated) probability density above, standard deviation in momentum and position are formally evaluated as
\begin{eqnarray}
	(\sigma_{k_x})^ 2 &=& \int_{_{-\infty}}^{^{+\infty}}\hspace{-.5 cm}dx\,\int_{_{-\infty}}^{^{+\infty}}\hspace{-.5 cm}dk_x \,k_x^ 2 \Tr[\mathcal{W} \gamma_0] = \frac{z^2 +1 }{2(1-z)}, \\
	(\sigma_{x})^2 &=& \int_{_{-\infty}}^{^{+\infty}}\hspace{-.5 cm}dx\,\int_{_{-\infty}}^{^{+\infty}}\hspace{-.5 cm}dk_x \,s^ 2\Tr[\mathcal{W} \gamma_0] = \frac{z^2 +1 }{2(1-z)},
\end{eqnarray}
exhibiting coincident values. For $z=0$, the uncertainty relation is saturated with $\sigma_{k_x} \, \sigma_{x} = 1/2$. On the other hand, it is unbounded for increasing values of the mixing parameter, approaching infinity for the maximally mixed and delocalized state with $z=1$, as it can be inferred from Fig.~\eqref{gaussianphasespace}.

\subsubsection{Quantum and classical information in Dirac-Gaussian states}
The information content of mixed Gaussian states is also relevant. In fact, using the change of variables as discussed above (cf. \eqref{gaussianintegration}), one now calculates the entropies with respect to both Hilbert spaces. Recalling that the phase-space entropy is evaluated by integrating the probability density squared (cf. (\ref{entropyps})), the result reads

\begin{eqnarray}
	\mathcal{I}_{\{x,k_x\}} &=& 1- 4\pi^2 \int_{_{0}}^{^{+\infty}}\hspace{-.5 cm}dr \, r \,\Tr[\mathcal{W} \gamma_0] ^2 \nonumber\\
&=&
	\frac{z}{2}\bigg(2 + z - z^2\bigg),
	\label{entropy1}
\end{eqnarray}
where the calculation is also left to Appendix \ref{AppC}. The minimum uncertainty state is obtained for $z=0$, which has zero entropy, whereas the completely delocalized Gaussian state with $z=1$ saturates the linear entropy at unity. Therefore, delocalization, with an arbitrarily small probability density in phase space, yields the maximum information content (or entropy) in this Hilbert space. 

Moving to the computation of the spin-parity entropy (cf. Appendix \ref{AppC}), the result can be written as
\begin{eqnarray}
	\mathcal{I}^{SP} &=& 1 -4\pi^2\Tr\left[\left(\gamma_{0} \,\int_{_{0}}^{^{+\infty}}\hspace{-.5 cm}dr \, r \,\mathcal{W} \right)^2\right] \nonumber \\
	 &=& \frac{1}{2}\left(1 + z - \frac{z^2}{2}\right),\label{entropy2}
\end{eqnarray}
which is finite for all Gaussian states, including the ground state. 

Wrapping up, the total mutual information is then composed by all three quantifiers above and reads
\begin{equation}
 M^{SP}_{\{x,k_x\}} = \frac{3z^2 + z - z^4}{2 + 2 z},
 \label{mutualmixed}
\end{equation}
vanishing only for $z=0$ in the region $0 \leq z \leq 1$, which confirms that the completely mixed state is not trivial.

This result is confronted with the expression for the quantum concurrence, 
\begin{eqnarray}
 \mathcal{C}^2 &=& \sum_{i=0} ^\infty p_i \mathcal{C}^2_i = p_0 \mathcal{C}^2_0 +\sum_{i\neq0} ^\infty p_i \mathcal{C}^2_i \nonumber \\
 &=& 0 + \sum_{i=1} ^\infty p_i\mathcal{C}^2_i \leq \frac{1-z}{2}\sum_{i=1} ^\infty z^i \nonumber \\ 
 &=& \frac{z}{2},
 \label{maximalconcurrence}
\end{eqnarray}
where the upper bound $\mathcal{C}^2_i = \frac{1}{2}$ is only valid in strong magnetic fields. Thus, $\mathcal{C}^2 = z/2$ is the greatest value for the intrinsic concurrence; otherwise, it cannot be calculated analytically. In any case, it minimizes the mutual information quantifier. To see this, the mutual information in Eq.~\eqref{mutualmixed} is expanded around the ground state, $z\approx0$, 
\begin{equation}
 M^{SP}_{\{x,k_x\}} = \frac{z}{2} + \mathcal{O}(z^2),
\end{equation}
concurring with the result just obtained. Thus, both the mutual information and the intrinsic concurrence increase linearly for small values of the mixing parameter. For greater values of $z$, classical correlations become relevant. 

\begin{figure} [H]
 \centering
 \includegraphics[width=0.6\textwidth]{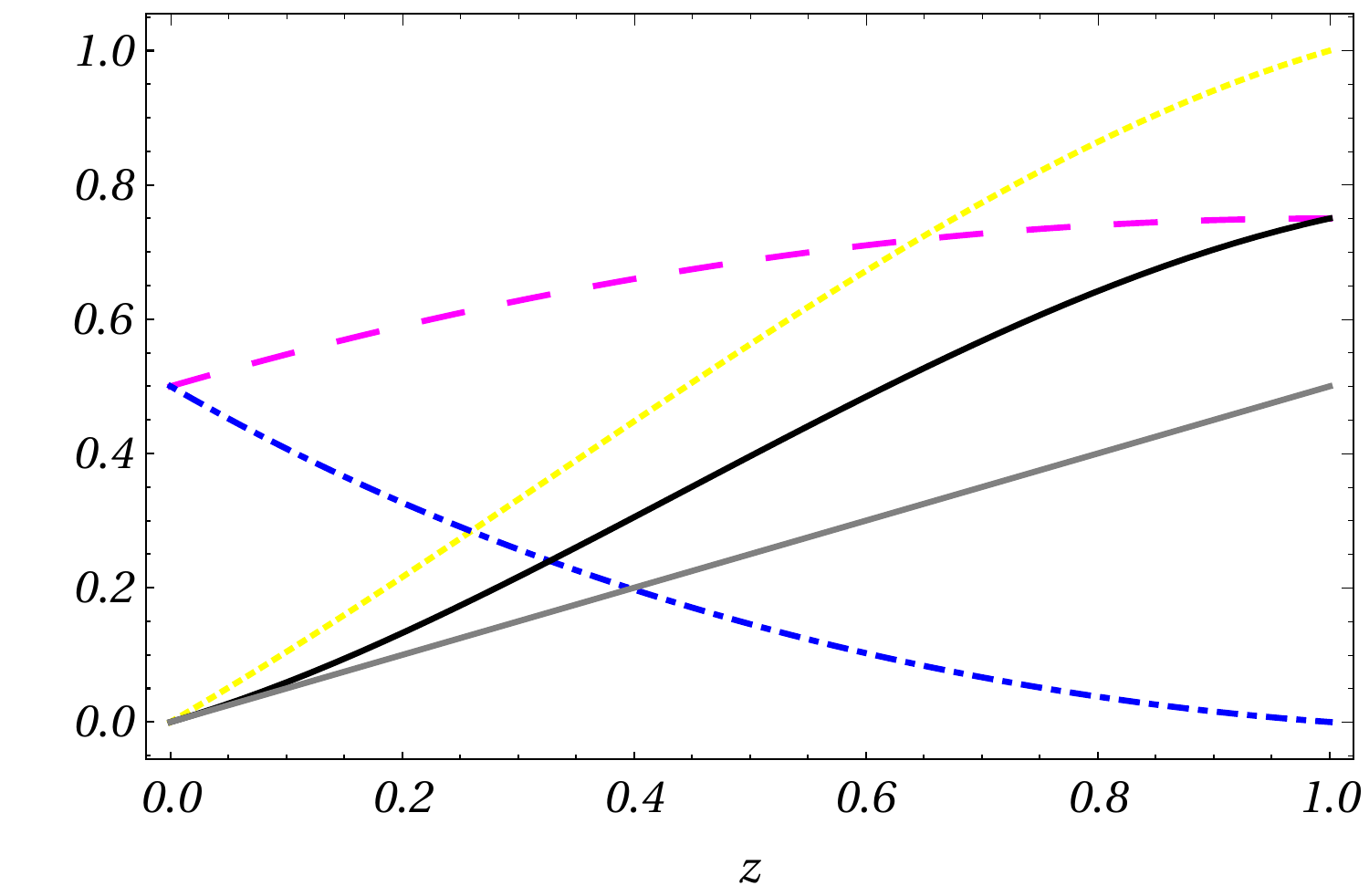}
 \caption{(Color online) Information balance for mixed Gaussian states. Linear entropies, $\mathcal{I}_{\{x,k_x\}} $ (cf. Eq.~\eqref{entropy1}; dotted yellow line) and $\mathcal{I}^{SP}$(cf. Eq.~\eqref{entropy2}; dashed magenta line), with respect to phase-space and spin-parity Hilbert spaces are depicted alongside the estimated concurrence (cf. Eq.~\eqref{maximalconcurrence}; solid gray line), and quantum purity (cf. Eq.~\eqref{puritymixed}; dot-dashed blue line). The total mutual information considering all these quantifiers (cf. Eq.~\eqref{mutualmixed}; solid black line) is also included. It is equal to the spin-parity entropy for a maximally mixed state, with $z=1$, which has zero purity and maximum phase-space entropy. }
 \label{fig2}
\end{figure}

These results are illustrated in Fig.~\eqref{fig2}, where all information quantifiers are depicted. The maximally mixed state, with $z=1$, indeed maximizes the phase-space linear entropy, $\mathcal{I}_{\{x,k_x\}} $, and is completely delocalized. Thus, for highly mixed states, the global density matrix loses the exact amount of information that is gained by the continuous-variable density matrix. In the aftermath of maximal mixing, the mutual information between phase and spin-parity spaces is solely described by discrete degrees of freedom. 

To sum up, the information profile of stationary Gaussian states has been examined in the phase space. It is somewhat curious that entropy measures only depend on the probability distribution due to randomness effects. The so-called coherent state has a vanishing phase-space entropy since it covers the minimum area, whereas the fully delocalized Gaussian state maximizes the same quantifier. In regards to the mutual information, the main result is that as the ground state contributes less to the mixture, this measure departs from the quantum concurrence and includes other correlations, simplifying to the spin-parity entropy at the maximally mixed state.

In the next section, one finally links mixed density matrices to thermalized low-dimensional electrons. The quantum information quantifiers are computed in terms of the relevant physical observable, namely, the partition function.

\section{Thermodynamic ensemble and high-temperature correlations}

The results from previous sections can be read for statistical ensembles as linear combinations of Wigner functions. Frequently, physical systems exhibit a probability distribution that is dictated by the qubit-environment interaction; in particular, the environment itself probabilistically selects the density matrix.
One now assumes that the probability is distributed according to energy eigenvalues. If the probability of measuring each state is canonically distributed \cite{Gibbs}, then an ensemble of Wigner functions describes 2-$dim$ thermalized systems. To see this, the energy eigenvalues are written as
\begin{equation}
	E_n= \sqrt{ \Delta^2 + 2n\,e\mathcal{B}} \, \, \mbox{with} \, \,\Delta^2 = m^2 + k_z ^2,
\end{equation}
and one possible statistical mixture is identified from the basis of equal-time Wigner functions as
\begin{equation}
 \mathrm{W}(s,\,k_x) = \sum_{n=0}^{\infty} p_n \left( \omega ^- _{n,1} \right) ,
 \label{ensemble1}
\end{equation}
where one assumes the unnormalized weights $p_n = \exp[ - \beta E_n]$ with $\beta = \frac{1}{ T}$, where $k_B=1$ is the Boltzmann constant and $T$ the equilibrium temperature\footnote{Dimensionless variables are discussed in Appendix \ref{AppD}. In natural units, both $\beta$ and $T$ are dimensionless.}. It has been assumed that for each Landau level, including the ground state, the quantum state is pure as it was introduced in Eq.~\eqref{9997}. One notices that the ground state solution does not have the two-fold spin degeneracy of excited Landau levels, justifying the ensemble with positive intrinsic parity $r=1$ and spin-down states only.

Precisely, this particular ensemble is somewhat arbitrary. Each Landau level can be either mixed or in a superposition state that affects the correlation profile as initially discussed in Fig.~\eqref{fig1}. A mixture within the same Landau level is ruled out since the statistical weights $p_n$ would have to be normalized again. However, a change of basis is also admitted, and one must investigate if the total information can be evaluated unambiguously. This multi-valuedness is expected since there are multiple quantum systems that are described by the same partition function.
\subsection{Partition function}
For all systems that can be thus identified, it is expected that the relevant observables of the theory are independent of the gauge-dependent Wigner functions chosen to realize this ensemble. Explicitly, one notices that Eq.~\eqref{ensemble1} is still not normalized, since
\begin{equation}\label{partition}
 \int_{_{-\infty}}^{^{+\infty}}\hspace{-.5 cm}dx\,\int_{_{-\infty}}^{^{+\infty}}\hspace{-.5 cm}dk_x \,\Tr[\mathrm{W} \gamma_0] = \sum_{n=0}^{\infty} \exp[ - \beta E_n] = \mathcal{Z},
\end{equation}
where the notation for the partition function $\mathcal{Z}$ is used.

Of course, for fermions in 3-d, the sum over states should take into account the quantum number related to the $k_z$ degree of freedom. Otherwise, Eq.~\eqref{partition} is the relevant object for the thermodynamic quantities of low-dimensional Dirac-like systems as, for instance, the graphene electrons with quantum numbers read as fixed parameters of the theory \cite{Houca}.
In fact, for graphene described as Dirac-like systems, the $\Delta$ parameter and the Rashba coupling play a similar role in the calculation of the partition function \cite{Houca}. Without this coupling, the partition function is formally obtained with $\Delta = 0$, however, with results constrained to the infinite-temperature limit ($T \to \infty$). To address this issue, the calculation will be revisited in the following. 

Let
\begin{equation}\label{eq1a}
	\mathcal{Z} = \sum_{n=0}^{\infty} \exp\bigg[- \mu \sqrt{\kappa + n} \bigg],
\end{equation}
with
\begin{equation}\label{thermalparameters}
	\mu = \frac{\sqrt{2 \,e\mathcal{B}}}{ T} \, \, \mbox{and} \, \, \kappa = \frac{\Delta^2}{2 \,e\mathcal{B}}.
\end{equation}
Now, the exponential can be expressed in terms of a complex contour integral, 
\begin{equation}\label{exp}
	e^{-z} = \frac{1}{2\pi i} \oint \,ds\, z^{-s} \Gamma(s) = \sum_{n=0} ^{+\infty} \frac{(-z)^n}{n!},
\end{equation}
where the integration is performed along a counterclockwise contour that includes all the poles of the Gamma function. A straightforward choice is the circle of radius $R\rightarrow \infty$ around the origin in the complex plane. To obtain the second equality with the residue theorem, one notices that the Gamma function $ \Gamma(s) $ can be defined in the complex plane as a meromorphic function with simple poles at nonpositive integers through the fundamental relation
\begin{equation}\label{gamma}
	\Gamma(s) = \frac{\Gamma(s+1)}{s},
\end{equation}
for any $s$ in the complex plane except at the poles \cite{Artin}. For positive integers, the Gamma function admits the usual factorial representation, $\Gamma(s) = (s-1)!$. The simple poles at $s=-n$, with $n$ a nonnegative integer, have residues given by
\begin{equation}\label{gammapoles}
\Res\big(\Gamma(s),-n\big) = \lim_{s\rightarrow -n} (s+n) \Gamma(s) = \lim_{s\rightarrow -n} \frac{ (s+n) \Gamma(s+ n+1)}{s(s+1)...(s+n)} = \frac{(-1)^n}{n!},
\end{equation}
where the recurrence relation (\ref{gamma}) has been used $n+1$ times to relate $\Gamma(-n)$ to $\Gamma(1)=1$. These residues can be plugged into Eq.~\eqref{exp} to finally obtain the second equality, which shows that the complex integral above is a valid representation of the exponential function, as long as all the residues are taken into account.

Turning back to the computation of the partition function above, the insertion of the integral into Eq.~\eqref{eq1a} yields
\begin{eqnarray}\label{eq1b}
	\mathcal{Z} &=& \frac{1}{2\pi i} \sum_{n=0} ^{+\infty} \oint \,ds\, \mu^{-s} (n+ \kappa)^{-s/2} \Gamma(s) \nonumber \\ 
&=& \frac{1}{2\pi i} \oint\,ds\, \zeta(s/2,\kappa) \,\Gamma(s) \, \mu^{-s},
\end{eqnarray}
where the series in the first line was written as the Hurwitz zeta function $\zeta(z,\kappa)$, which simplifies to the well-known Riemann zeta function when $\kappa =1$, $\zeta(z,1) \equiv \zeta(z)$ \cite{Hasse}. The series converges when $\Re(z) > 1$ and is analytic continued otherwise, namely, to all $z$ in the complex plane and $\kappa > 0$ except at the simple pole $z=1$.\footnote{In \cite{Hasse}, the Hurwitz zeta function is defined as $\zeta(s,w) = \sum_{n=1} ^\infty \frac{1}{(w+n)^s}$, and $w = \kappa -1$ in the current notation. Thus, $w>-1$ implies $\kappa >0$. } The continuation is necessary since the residues of the integrand need to be evaluated at the poles of the Gamma function (cf. (\ref{gammapoles})). Moreover, since the integration contour contains the real axis, the new singularity is already included.\footnote{The series representation of zeta functions diverges in the half plane $\Re(z) \leq 1$. Therefore, the uniqueness of the analytic continuation allows one to extract the relevant residue, and an adequate contour should include all the singularities, which profoundly affect the calculation. This assumption will be tested in the following so as to confirm if the relevant quantifiers are physically acceptable.} The residue theorem is once again applied to $z=1$,
\begin{equation}\label{residue}
\Res\big(\zeta(z,\kappa), 1\big) =	\lim_{z \rightarrow 1} \, (z-1) \, \zeta(z,\kappa) = 1,
\end{equation}
where the result is obtained by considering an alternative representation of the zeta function (cf. Appendix \ref{AppE}).

Finally, the partition function is obtained by including the residues of the poles at nonpositive integers and also at $s=2$. The final expression for the partition function reads
\begin{equation}\label{eq1c}
\mathcal{Z} = \sum_{m=0} ^{+\infty} \zeta\left(-\frac{m}{2}, \kappa\right) \frac{(-\mu)^m}{m!} + \frac{2}{\mu^2}.
\end{equation}
The series is then regarded as an expansion on the parameter $\mu = \frac{\sqrt{2 \,e\mathcal{B}}}{T}$, which is nowhere obvious from the formal definition of the partition function. It is absolutely convergent, due to the analytic continuation of zeta functions, and determines the low temperature regime, whereas the last term in Eq.~\eqref{eq1c} dominates for high temperatures. This is consistent with \cite{Santos}, where the expansion and the leading term as $\mu \approx 0$ are both present. 

In the current framework, correlations are investigated for small deviations from the infinite-temperature limit, and a few terms of the series are included in computations. One notices that the two representations of the partition function, (\ref{eq1a}) or (\ref{eq1c}), are useful in different temperature ranges. In the former case, it simplifies to a single term at absolute zero; in the latter one, it reduces to a single term in the infinite-temperature limit. 
\subsubsection{Thermodynamic functions}
\begin{figure}
	\centering
	\includegraphics[width=0.8\textwidth]{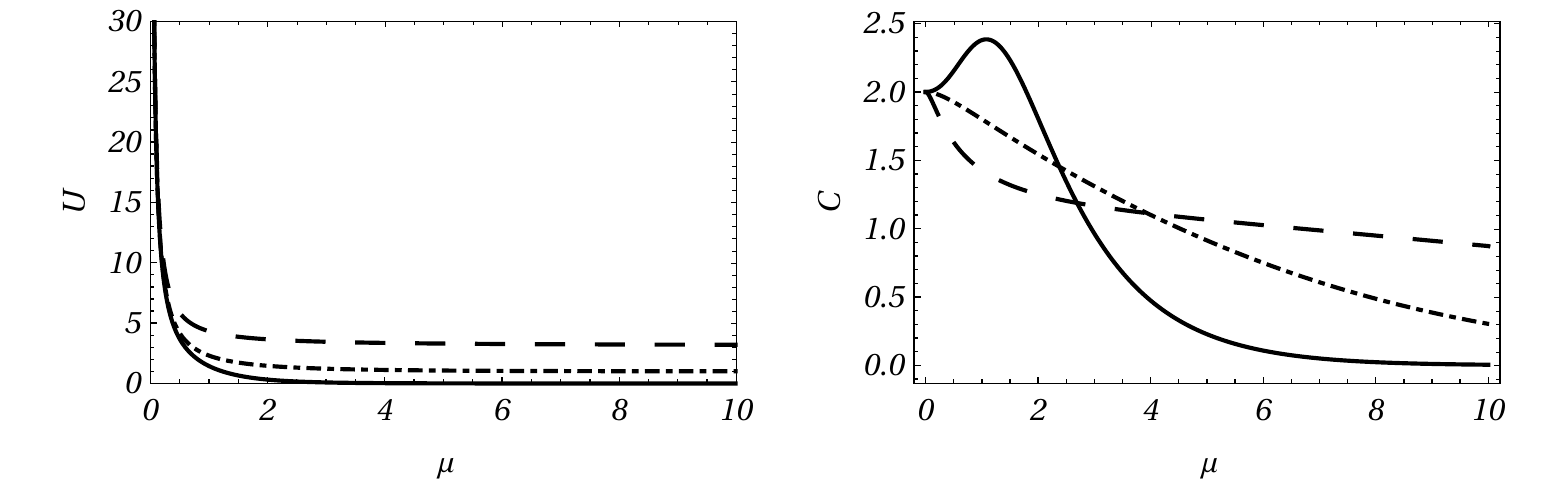}
	\caption{Dimensionless internal energy (left plot) and specific heat (right plot) in high temperatures.  In natural units, $	\mu = \frac{\sqrt{2 \,e\mathcal{B}}}{ T}$ is the expansion parameter for the partition function. Results are for $\kappa = 0$ (solid line), $1$ (dot-dashed line), and $10$ (dashed line) when some few terms from the infinite expansion (cf. Eqs.~\eqref{eq1a} and \eqref{exp}) are considered.
At $\mu=0$, the result is analytic and independent of $\kappa = \frac{m^2+ k_z ^2}{2 \,e\mathcal{B}}$ (cf. Eq.~\eqref{thermalparameters}).}\label{thermofunctions}
\end{figure}

For completeness, usual dimensionless thermodynamic functions are redefined (cf. Appendix \ref{AppD}) as, for instance, 
\begin{eqnarray}
	U &=& -\sqrt{2\,e\mathcal{B} } \, \frac{\partial \ln (\mathcal{Z})}{\partial \beta} = - \frac{\partial \ln (\mathcal{Z})}{\partial \mu} \nonumber \\
	&=& \frac{-1}{\mathcal{Z}} \left\{ \sum_{m=0} ^{+\infty} \zeta\left(-\frac{m+1}{2}, \kappa\right) \frac{(-\mu)^m}{m!} - \frac{4}{\mu^3}\right\},
\end{eqnarray}
the internal energy, and
\begin{eqnarray}
	C &=& - \mu^2 \left(\frac{\partial U}{\partial \mu}\right) \nonumber \\
	&=& -\mu^2 \left\{ U^2 - \frac{1}{\mathcal{Z}} \left(\sum_{m=0} ^{+\infty} \zeta\left(-\frac{m+2}{2}, \kappa\right) \frac{(-\mu)^m}{m!} + \frac{12}{\mu^4}\right) \right\},
\end{eqnarray}
the specific heat. They are depicted in Fig.~\eqref{thermofunctions}, which includes the low-temperature corrections. On the one hand, results are not very illuminating as $\kappa$ increases indefinitely, since more terms are needed to obtain the adequate precision. On the other hand, the infinite-temperature limit is independent of the $\kappa$ parameter. In fact, at $\mu=0$, the Dulong-Petit law is retrieved for the specific heat in two dimensions. This could be anticipated, since the partition function takes the same form as, for instance, for graphene electrons close to Dirac points \cite{Santos}. 

\subsection{Temperature-dependent maximal mixing}

Turning back to the computation of the relevant quantum information observables for finite temperatures, the normalized Wigner function is recast as $\mathcal{W}_{TE} =(\mathcal{Z})^{-1} \mathrm{W}$, so as to ensure the probability is conserved.

\begin{figure}
	\centering
	\includegraphics[width=0.4\textwidth]{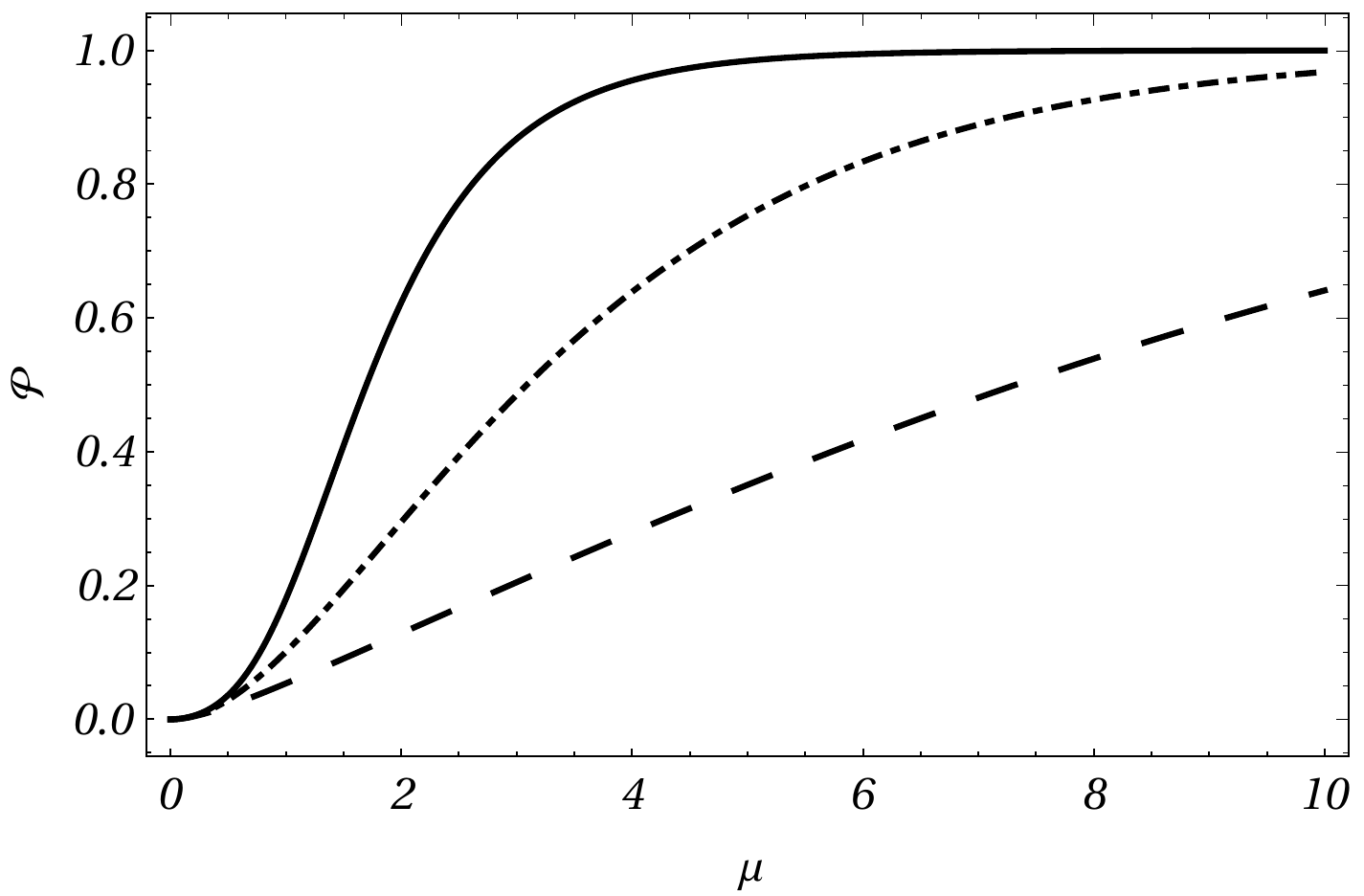}
	\caption{Quantum purity of uneven Landau levels in finite temperatures. As in Fig.~\eqref{thermofunctions}, plots are for $\kappa = 0$ (solid line), $1$ (dot-dashed line), and $10$ (dashed line). At absolute zero ($T \to 0$), $\mu \rightarrow \infty$, and all ensembles approach asymptomatically $\mathcal{P} = 1$, a pure state. In the infinite-temperature limit ($T \to \infty$), $\mu = 0$, all states are maximally mixed. }\label{thermalpurity}
\end{figure}

Using Eq.~\eqref{puritydem}, the quantum purity can be evaluated, 
\begin{eqnarray}
 \mathcal{P}[\mathcal{W}_{TE}] &=& \mathcal{Z}^{-2} (\mu) \sum_{n\neq0}^{\infty} \left( \exp[ - \mu \sqrt{\kappa + n}] \right) ^2 + \nonumber \\
 &=& \mathcal{Z}^{-2} (\mu) \left( \sum_{n=0}^{\infty} \exp[ - 2 \mu \sqrt{\kappa + n}] \right) \nonumber \\
 &=& \mathcal{Z} ^{-2} (\mu) \, \mathcal{Z} (2 \mu) \\
 &\approx& \frac{\mu^2}{8},\label{puritythermal} 
\end{eqnarray}
where the last line is valid in the high-temperature limit (cf. Eq.~\eqref{eq1c}). Thus, the quantum purity depends only on the partition function, evaluated at two distinct temperatures, $\mu$ and $2 \mu$. This result is in fact consistent with \cite{BernardiniPHYSA}, where the quantum purity has been universally derived from a partition function. At absolute zero ($T \to 0$), however, the quantum purity is more easily obtained from Eq.~\eqref{eq1a}, whose leading term yields
\begin{equation}
 \mathcal{P}[\mathcal{W}_{TE}] _ {T=0} = \lim_{\mu \rightarrow \infty}	 \frac{\mathcal{Z}(2\mu)}{\mathcal{Z}(\mu)^2} = \lim_{\mu \rightarrow \infty} \frac{\exp[-2\mu \kappa^{1/2}]}{\exp[-\mu\, \kappa^{1/2}]^2} = 1,
\end{equation}
i.e. the quantum system is pure and is in the ground state for any value of $\kappa$, as expected. This concurs with the numerical results exhibited in Fig.~\eqref{thermalpurity}, which shows that the quantum purity is indeed constrained to $0 \leq \mathcal{P} \leq 1$. It is somewhat surprising that the highly oscillatory behavior of zeta functions never violates the admissible values of quantum purity. This behavior is obtained when an ever increasing number of terms is included at low temperatures.

This result is reassuring since the continuation of zeta functions to negative integers still yields a consistent quantum purity quantifier. Moreover, Fig.~\eqref{thermalpurity} shows that purity in thermodynamic ensembles for uneven Landau levels has a straightforward interpretation. Starting at absolute zero, the plotted curves indicate that the system is in the ground state. As the temperature increases, the system also occupies excited levels. For small values of $\kappa$, the lowest energy is away from the first excited level, and the system remains in the ground state up to great temperatures. Conversely, for increasing values of $\kappa$, the zero-point energy level is much closer to excited levels, and thus only a tiny amount of thermal energy is required to populate excited levels; namely, for $\kappa \rightarrow \infty$, the system remains in the ground state strictly at absolute zero. For $\kappa = 0,1$, the system remains in the ground state up to $\mu \approx 5,10$, respectively. This is characteristic of uneven energy levels. In contrast, an energy spectrum with constant energy spacing can be simply shifted by a constant amount. Thus, the quantum purity would not depend on any additional parameter.

To summarize, the partition function has been computed using the method of complex contour integral. Even though the low-temperature behavior does not have a closed form expression, the quantum purity quantifier, which depends only on the partition function itself, ranges from a maximal mixture in the infinite-temperature limit to a pure state at absolute zero. It is emphasized that the quantum purity is straightforwardly computed once the partition function is obtained. Next, this result will be used to finally compute the temperature-dependent correlations for Dirac spinors. 

\subsection{Intrinsic correlations}\label{subsecb}
Unlike the quantum purity expression, which solely depends on the probability distribution, other entropy measures do not have an obvious dependence on the partition function. Even worse, the energy parameters of each Wigner function from Eq.~\eqref{9999} make sums difficult to obtain because one needs to consider the spinor degrees of freedom. Nevertheless, it is essential to verify if a general behavior is observed in the infinite-temperature limit. In particular, correlations are expected to exhibit temperature-dependent plateaus, i.e. a temperature for which these quantifiers become saturated. They can be evaluated even if the density matrix does not exhibit a closed form because orthogonality in phase space can be used to calculate phase-space integrals.
 
In the following computations, it will be assumed that $\kappa = 0$ (cf. (\ref{thermalparameters})), valid in strong magnetic fields. The associated energy parameters take the form 
\begin{eqnarray}
 B_n &=& \frac{\sqrt{2n\,e\mathcal{B}}}{\sqrt{m^2 + k_z^2 + 2n\,e\mathcal{B}} +m} = 1-\delta_{n,0}, \\
 A_{n \neq 0} &=& \frac{k_z}{\sqrt{m^2 + k_z^2 + 2n\,e\mathcal{B}} +m} = 0, \\
 \eta_{n \neq 0}&=&\frac{\sqrt{m^2 + k_z^2 + 2n\,e\mathcal{B}} +m}{2\sqrt{m^2 + k_z^2 + 2n\,e\mathcal{B}}} = \frac{1}{2}.
\end{eqnarray}
In other words, for $\kappa =0$, the constants above take an universal value for excited levels. However, the normalization of the ground state is still undetermined but is constrained to $1/2\leq\eta_0\leq1$. 

The reason for this assumption is threefold. First, the correlation structure no longer depends on the particular Landau level,\footnote{This is true only after phase-space averaging. The local behavior of the Wigner function is still quantized and described by the HO basis.} simplifying summations over several indices. Indeed, from the preliminary results shown in Fig.~\eqref{fig1}, the total information also takes a constant value. Second, the partition function at $\kappa =0$ describes thermalized 2-$dim$ Dirac electrons on graphene \cite{Santos}. Third, one is mainly interested in the behavior of correlations in the infinite-temperature limit, which is expected to be independent of $\kappa$.

Now, recalling that the ensemble from Eq.~\eqref{ensemble1} is given in terms of $	\omega^-_{n,1}(s,\,k_x)$ introduced in Eq.~(\ref{9997}), which for $n > 0$ and $\kappa = 0$ simplifies to 
\begin{eqnarray}
	\omega^-_{n,1}(s,\,k_x) &=& \frac{1}{2} \left( \begin{array}{cccc} 
		0 & 0& 0& 0\\
		0 & \mathcal{L}_{n} & \mathcal{M}_{n}& 0 \\ 
		0 & -\,\mathcal{M}_{n} & -\,\mathcal{L}_{n-1}& 0 \\ 
		0 & 0 & 0& 0 \\ 
	\end{array} \right).
\end{eqnarray}
The ground state needs an additional normalization, which does not qualitatively affect the result for high temperatures. The phase-space entropy is calculated for $\mathcal{W}_{TE}$ (cf. Appendix \ref{AppF}),
\begin{equation}\label{ikxthermal}
 \mathcal{I}_{\{x,k_x\}} = 1 - \frac{\mathcal{Z}(2\mu)}{2\mathcal{Z}^2 (\mu) } - \frac{p_0 ^2}{2\mathcal{Z}^2} - \frac{1}{2 \mathcal{Z}^2 } \sum_{n = 0} p_n \, p_{n+1}.
\end{equation}
It has been noticed that for high temperatures, the third and fourth terms can be dropped.

Therefore, the phase-space linear entropy is approximated as $ \mathcal{I}_{\{x,k_x\}} \approx 1 - 1/2 \, \mathcal{P}$, approaching unity at $\mu = 0$. A similar result has been obtained for Gaussian states in the previous section. The quantum state is spread out in phase space, and thus any phase-space measurement yields the maximal amount of information with respect to the continuous degrees of freedom. Conversely, at absolute zero, Eq.~\eqref{ikxthermal} is evaluated with $\mathcal{Z} = p_0 = 1$, and thus $\mathcal{I}_{\{x,k_x\}} = 0$, i.e. the ground state has zero entropy. 

\begin{figure}
	\centering
	\includegraphics[width=0.4\textwidth]{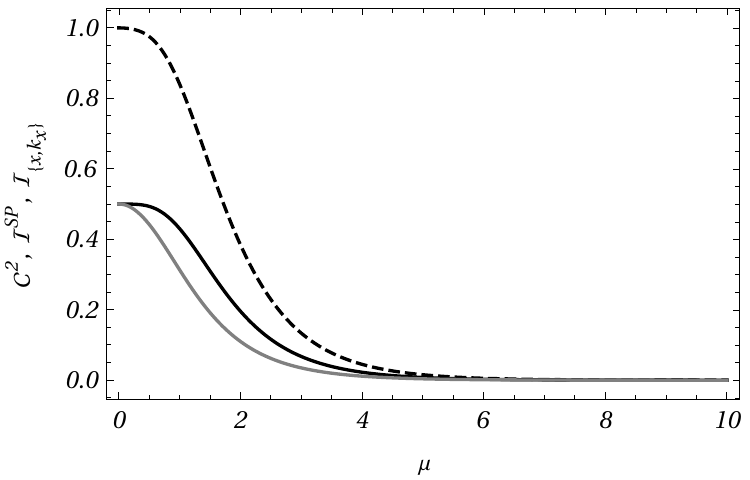}
	\caption{Finite-temperature information profile of uneven Landau levels. It is assumed that $\kappa=0$ (cf. (\ref{thermalparameters})), which is valid in strong magnetic fields. In the infinite-temperature limit, with $\mu=0$, these results do not depend on $\kappa$. Solid and dashed black lines are for spin-parity (cf. Eq.~\eqref{ispthermal}) and phase-space (cf. Eq.~\eqref{ikxthermal}) entropies, respectively. Solid gray line is for the intrinsic concurrence (cf. Eq.~\eqref{concthermal}). The result shows that these entropy measures increase for increasing temperatures, and quantum correlations are preserved. }\label{entropythermal}
\end{figure}

The result obtained for the quantum purity is closely tied with the continuous linear entropy, which can be confirmed in Fig.~\eqref{entropythermal} since the linear entropy is, to a good approximation, a reflection of the quantum purity curve. Qualitatively, these quantifiers have an opposite behavior for finite temperatures. For $\kappa = 0$, the quantum state has zero entropy when $\mu \gtrsim 5$, the temperature below which the quantum state is pure. This result can be generalized to different types of interactions as long as the particle is confined. A maximally mixed state corresponds to an arbitrary statistical mixture of infinitely many eigenstates of a given Hamiltonian. It follows that the phase-space entropy quantifier is always maximized, since the basis is orthogonal, and thus mixed states cover the entire phase space. 

The calculation of the spin-parity linear entropy follows along the same lines and is left to Appendix \ref{AppF} for brevity. The final result for $\kappa=0$ is
\begin{equation}\label{ispthermal}
 \mathcal{I}^{SP} = \frac{1}{2}\left( 1 - \frac{p_0^2}{\mathcal{Z}^2}\right) ,
\end{equation}
where it has been assumed the same the ground state normalization from Eq.~\eqref{ikxthermal}. One notices that the information associated with the discrete degrees of freedom is also enhanced by greater temperatures. 

Finally, one calculates the concurrence as
\begin{eqnarray}\label{concthermal}
	\mathcal{C}^2 _{TE} &=& \mathcal{Z}^{-1}\sum_{n=0} p_n \mathcal{C}^2 _n = \mathcal{Z}^{-1}\sum_{n=1} 2 p_n (\eta_n B_n)^2 \nonumber \\
	&\leq& \frac{1}{2\mathcal{Z}} \left( \mathcal{Z} - p_0 \right),
\end{eqnarray}
where the equality holds when $\kappa=0$, the case under consideration. $p_0$ shows up again because the ground state has zero intrinsic concurrence. One briefly recalls that this measure is affected by the basis chosen, as it has been preliminarily discussed in Fig.~\eqref{fig1}. Indeed, there are ensembles with the same partition function that only exhibit classical correlations. 

Results for the partial entropies and concurrence are summarized in Fig.~\eqref{entropythermal}, confirming what is suggested by Fig.~\eqref{thermalpurity}. For increasing temperatures, the decrease of quantum purity compensates the increase of phase-space entropy. This compensation effect is exact in the infinite-temperature limit, where the quantum purity and the continuous-variable entropy reach $0$ and $1$, respectively. Thus, the spin-parity entropy becomes the only remaining source of information, and $	M^{SP}_{\{x,k_x\}} = \mathcal{I}^{SP} = \mathcal{C}^2 _{TE}$ at $\mu = 0$ (cf. (\ref{mutualinfo})). The total mutual information is not affected by choosing $\kappa \neq 0$ in the infinite temperature limit. However, quantum correlations decrease by increasing $\kappa$ or choosing a distinct superposition.

In brief, partial entropies have been computed analytically for finite temperatures in terms of the partition function. One has also measured quantum correlations with the concurrence quantifier. The amount of entanglement saturates due to highly excited Landau levels as the temperature increases. In the infinite-temperature limit, the quantum state spreads over phase space, and thus maximizes the phase-space linear entropy while minimizing the quantum purity measure. 

\section{Conclusions and Outlook}
One has investigated stationary mixtures of Dirac-like Wigner functions that carry the intrinsic correlation structure as $SU(2) \otimes SU(2)$ qubits in relativistic Landau levels \cite{BernardiniEPJP}. The preliminary result in the present study is the uniqueness of pure state decompositions of Wigner functions in some kind of a confining potential. The external magnetic field introduces a new degree of freedom, quantizing not only the energy spectrum, but also the density matrix: a result that simplifies the calculation of the quantum concurrence. This is completely consistent with the mutual information measure, which quantifies the total information shared by the phase-space and spin-parity (Hilbert) spaces. 

As a prelude, superpositions and mixtures have been contrasted. The fundamental difference is summarized as follows. The interference between two solutions in configuration space degrades the intrinsic entanglement. However, probability distributions in phase space do not affect this separability measure. Only classical correlations are affected by mixedness.

One has then considered mixed Gaussian states analytically obtained from the generating function of Laguerre polynomials. The physical observables only depend on the mixing parameter, which works as a localization probe in phase space. In fact, the ground state has zero phase-space entropy and is the most localized state as allowed by quantum mechanics. On the other extreme, the fully delocalized state covers the entire phase space, has zero purity but maximum phase-space entropy.

The phase-space linear entropy is maximized when the quantum purity is minimized. Thus, this entropy quantifier becomes a localization measure in its own right. Since all the remaining information, quantum and classical, is quantified by the spin-parity linear entropy, one concludes that it contains the quantum concurrence (squared). Likewise, the non-separability measure between discrete degrees of freedom is thus regarded as a source of uncertainty, or (quantum) information, in this Hilbert space. 

Some conclusions drawn for closed-form Gaussian states are not restricted to these particular states. A similar behavior is also valid for any confining potential, which depends on position variables. Namely, all completely mixed Wigner functions maximize the phase-space linear entropy because systems with infinite-dimensional bases need a small but finite probability over all eigenstates to reach a zero purity state. For orthogonal eigenstates, maximally mixed states are thus delocalized. 

Finally, one has investigated the correlation structure of canonically distributed probability distributions. One has obtained the partition function of the system with complex integration methods, which is the main result of the paper. It reads
\begin{equation*}
	\mathcal{Z} = \sum_{m=0} ^{+\infty} \zeta\left(-\frac{m}{2}, \kappa\right) \frac{(-\mu)^m}{m!} + \frac{2}{\mu^2}.
\end{equation*}
This partition function simplifies in the infinite-temperature limit to $\mathcal{Z} = 2/\mu^2$, independent of any additional parameter. However, the low-temperature corrections are described by analytically continued Hurwitz zeta functions $\zeta\left(-\frac{m}{2}, \kappa\right)$, which also include the dependence on the one-particle parameters $\kappa = \frac{m^2+k_z^2}{2 \,e\mathcal{B}}$. The quantum purity, $\mathcal{P} = \mathcal{Z}(2\mu)/\mathcal{Z}(\mu)^2$, has been immediately obtained. It quantifies the Landau levels occupation with respect to the fundamental state. Namely, at absolute zero, the system is in the ground state with $ \mathcal{P} = 1$. As the temperature increases, the quantum purity decreases down to $ \mathcal{P} = 0$, when all energy levels are occupied with an arbitrarily small probability.

In regards to entropy measures, a relevant behavior is observed in the infinite-temperature limit. Namely, the phase-space linear entropy is maximal when the state delocalizes, confirming the observation for Gaussian states. Thus, the spin-parity linear entropy becomes the only information source. This remaining information is either classical- or quantum-like, which in the latter case can also be quantified in terms of the quantum concurrence. However, a basis change affects the intrinsic entanglement. Each choice identifies a distinct quantum system.

Precisely, the partition function does not uniquely identify the (thermalized) low-dimensional system. For instance, graphene is a platform that appeals to the quantum information framework described here. One first recalls that graphene is a 2-$dim$ layer of carbon atoms ($A-B$) packed in a hexagonal lattice and whose low-energy excitations are described by Dirac Hamiltonians at the corners of the Brillouin zone. Briefly using the notation from Ref.~\cite{Goerbig}, the effective low-energy Hamiltonian is written as
\begin{equation}
H_{\bf{p}}=
 v_F \sigma^z \otimes \bf{p}\cdot {\boldsymbol \sigma},
\end{equation}
for small deviations $\bf{q}$ ($ \bf{p} = \hbar \bf{q}$) from the Dirac points, and $v_F$ the fermi velocity. Thus, the first Pauli matrix $\sigma^z$ denotes the valley pseudospin ($K-K'$), and the second set of Pauli matrices represents the sublattice pseudospin ($A-B$). Landau levels are formed when a perpendicular magnetic field is introduced with the minimal substitution, ${\bf p}\rightarrow{\bf p}-e{\bf A}$. Excited levels can be written in the basis\footnote{The Hamiltonian is written as a tensor product if the roles of the $A-B$ sublattices are reversed for the $K'$ valley.} $	\Psi = (\psi_{\bf{p},K}^A \, \, \psi_{\bf{p},K} ^B \, \,
\psi_{\bf{p},K'}^B \, \, \psi_{\bf{p},K'}^A)^T $ and are proportional to
\begin{equation}
	\Psi^\pm _{n,k_y}(\mathbf{x})= \exp(-ik_y y)\left(\begin{array}{c}
		\phi_{n-1}\\
		 \pm \phi_{n} \\
		 	\phi_{n-1} \\
		 (\mp)\phi_{n}
	\end{array}\right)
\end{equation}
where the upper sign corresponds to the conduction band, and the lower sign to the valence band, with eigenvalues $\epsilon_n = \pm \frac{\hbar v_F}{l_B} \sqrt{2n}$. The HO basis in configuration space is written as \cite{CastroNeto}
\begin{equation}
\phi_n = \displaystyle \frac{H_{n}\left( \displaystyle (x-l_B^2k_y)/l_B\right)}{\sqrt{2^{n}\,n!\sqrt{\pi}\,l_B}}\exp{\left[-\frac{1}{2} \left(\frac{x-l_B ^2k_y}{l_B}\right)^2\right]}, 
\end{equation}
where $l_B=\sqrt{ \hbar/eB}$ is the magnetic length and $H_n(x)$ is a Hermite polynomial. Thus, the scalar functions are closely related to those discussed for the $3$ spatial dimensions case\footnote{
The two copies of the massless $(2+1)$-$dim$ Dirac equation are not equivalent to the $(3+1)$-$dim$ Dirac equation and thus do not have a common set of eigensolutions. However, one notices that for a fixed Landau level, a map between these solutions is obtained by a straightforward eigenfunction expansion (cf. Appendix \ref{AppA}), 
\begin{eqnarray} 
	u^+_{n,1}(s) + u^-_{n,2}(s)&\propto & 	\Psi^+ _{n,k_y}(\mathbf{x}),\label{sup1}\\
-	u^-_{n,1}(s) + u^+ _{n,2}(s)&\propto& 	\Psi^- _{n,k_y}(\mathbf{x}), \label{sup2}
\end{eqnarray} 
(with $k_z = m =0$) up to a normalization factor. There is an additional $\exp(-ik_y y)$ factor at both sides that is factorized out.}. Thus, the stationary states of the low-dimensional Hamiltonian are expanded in the basis of the spin-parity Dirac Hamiltonian (an equivalent basis in $3$ spatial dimensions) if one simply identifies $v_F \leftrightarrow c$. This expansion involves only Landau levels with the same index. After this mapping, these Hamiltonians share the same eigenvalues, 
\begin{equation}
	\epsilon_n = \pm \frac{\hbar v_F}{l_B} \sqrt{2n} \leftrightarrow \pm E_n = \pm \frac{\hbar c}{l_B} \sqrt{2n},
\end{equation}
and thus the conduction band is associated to positive parity states, whereas the valence band is associated to negative parity states. 

The superposition of distinct parity states is not stationary, different from Eq.~\eqref{superposition} ($\sin (\theta) \, u^+ _{n,1}(s) + \cos (\theta) \, u^- _{n,1}(s)$). Nevertheless, at $t=0$ and $\theta = \pi/4$, all quantum information measures exhibit coincident values. This correspondence follows from the fact that this stationary state differs from Eqs.~\eqref{sup1}-\eqref{sup2} only by a relative sign in the spinor components.

Therefore, the proposed quantifiers are appropriate for measuring valley-sublattice correlations implied by the effective Hamiltonians close to Dirac points. From our results, when the energy eigenvalues are dominated by the magnetic field ($B_n =1$) in Fig.~\eqref{fig1}, the total mutual information reaches unity. However, the intrinsic concurrence between valley and sublattice reads (cf. Appendix \ref{AppB}) 
 \begin{equation}
 	\int_{_{-\infty}}^{^{+\infty}}\hspace{-.5 cm}ds\,\int_{_{-\infty}}^{^{+\infty}}\hspace{-.5 cm}dk_x \,	\mathcal{C}^2[\omega_{n,\theta}] = \frac{1}{2} \, \cos ^2 (2\theta)\vert_{\theta = \pi/4} = 0 ,
 \end{equation}
i.e. these qubit states are separable. This result could be anticipated since the Hamiltonian is diagonal in the valley subspace. To put it simply, the total information stored in these Dirac bispinors is composed by two parts: the entropy associated to continuous degrees of freedom, the HO basis, and the entropy associated to discrete degrees of freedom, the valley-sublattice pseudospins. However, these qubits are not entangled. One thus expects a finite concurrence measure if intervalley couplings are introduced.

In such context, the partition function obtained here describes the thermodynamics of graphene electrons close to Dirac points. Entropy measures have been obtained in terms of dimensionless variables that include a scaling factor, being consistent with both the relativistic and the solid-state Hamiltonians. Even though quantum correlations are trivial without any intervalley coupling, this framework is suitable not only to electronic band structures, but also to recently reported Dirac Hamiltonians for bosonic systems \cite{Kumar, Sun} with a confining potential.

\vspace{.5 cm}
{\em Acknowledgments -- The work of A. E. B. is supported by the Brazilian Agencies FAPESP (Grant No. 2023/00392-8) and CNPq (Grant No. 301485/2022-4). The work of C. F. S. is supported by the Brazilian Agency Capes (Grant No. 88887.499837/2020-00).}

\appendix 
\section{Equal-time Wigner function in Landau levels }
\label{AppA}
Considering that the prescription for calculating the equal-time Wigner function is indicated in Eq.~\eqref{equaltimewignerfunction}, the eigenfunctions of the Dirac Hamiltonian for a constant magnetic field can be written as 
\begin{equation}
	\psi = \exp\big[i((-1)^r E_n t + k_y y + k_z z)\big] u_{n,r} ^\pm (s_r), \label{stationarysolutions}
\end{equation}
describing plane-wave solutions in both $y$ and $z$ directions. The dynamics along the $x$-coordinate is shifted according to the charge of the particle,
\begin{equation}
	s_r = \sqrt{e {\mathcal B}} \left( x + (-1)^r\frac{ k_y }{ e{\mathcal B}}
	\right).
	\label{222}
\end{equation}
The bispinors $u_{n,r} ^\pm (s_r)$ form a complete basis in configuration space. It is possible to choose a set with positive parity states ($r=1$), 
\begin{eqnarray}
	u^+_{n,1}(s_1) = \sqrt{\eta_{n}}\left( \begin{array}{c} 
		\mathcal{F}_{n-1}(s_1) \\ 0 \\ 
		A_{n}\, \mathcal{F}_{n-1}(s_1) \\
		-B_{n}\, \mathcal{F}_{n} (s_1) 
	\end{array} \right), \quad 
	u^-_{n,1}(s_1) = \sqrt{\eta_{n}}\left( \begin{array}{c} 
		0 \\ \mathcal{F}_{n} (s_1) \\
		-B_{n}\,
		\mathcal{F}_{n-1}(s_1) \\ 
		-A_{n}\,\mathcal{F}_{n}(s_1)
	\end{array} \right), \quad 
\end{eqnarray}
and negative parity ($r=2$) states,
\begin{eqnarray}
	u^+_{n,2}(s_2) = \sqrt{\eta_{n}}\left( \begin{array}{c} 
		B_{n}\,
		\mathcal{F}_{n-1}(s_2) \\ 
		A_{n}\,\mathcal{F}_{n}(s_2) \\ 
		0 \\ \mathcal{F}_{n} (s_2)
	\end{array} \right), \qquad 
	u^-_{n,2}(s_2)= \sqrt{\eta_{n}}\left( \begin{array}{c} 
		-A_{n}\,
		\mathcal{F}_{n-1}(s_2) \\ 
		B_{n}\,
		\mathcal{F}_{n} (s_2) \\ 
		\mathcal{F}_{n-1}(s_2) \\ 0
	\end{array} \right),
\end{eqnarray}
For a clear connection with typical superposition phenomena, one simply chooses one of the two equivalent Hamiltonians for positively or negatively charged particle, fixing the $s_{r=1} \equiv s$ coordinate. The second set of solutions describes the usual $E<0$ solutions, and the HO basis in configuration space in terms of Hermite polynomials $ H_{n}(s)$ becomes
\begin{equation}
\mathcal{F}_{n}(s) = \left( \frac{ \sqrt{e{\mathcal B}}}{n! \, 2^n \sqrt{\pi}} \,
	\right)^{1/2} e^{-s^2/2} H_{n}(s)
\end{equation}

The corresponding matrix basis in Eqs.~(\ref{9996})-(\ref{9999}) can be obtained componentwise by plugging the Dirac bispinors above into Eq.~\eqref{equaltimewignerfunction}. Instead of calculating all $16$ components, one simply notices that there are only two distinct functions to be evaluated, 
\begin{eqnarray}
	\mathcal{L}_n (s,k_x) &=& \int_{_{-\infty}}^{^{+\infty}}\hspace{-.5 cm}du \,\exp[2 i\, k_x\, u]\,\mathcal{F}_{n}(s-u)\mathcal{F}_{n}(s+u) \\
		\mathcal{M}_{n} (s,k_x) &=& \frac{1}{2}\int_{_{-\infty}}^{^{+\infty}}\hspace{-.5 cm}du \,\exp[2 i\, k_x\, u]\,\bigg(\mathcal{F}_{n-1}(s-u)\mathcal{F}_{n}(s+u)+ \mathcal{F}_{n-1}(s+u)\mathcal{F}_{n}(s-u)\bigg),
\end{eqnarray}
which result in Laguerre polynomials as shown in Eqs.~\eqref{norr01} and \eqref{norr02}. These functions appear alongside the appropriate combination of constants $A_n$ and $B_n$.
\section{Pure state quantum concurrence}
\label{AppB}
For pure states, quantum correlations are quantified by the quantum concurrence whose formula for two qubits can be straightforwardly extended to the equal-time Wigner function, by recalling that the density matrix of the system $\varrho = \gamma_{0}\,\omega^{\pm}_{n,r}$ now also depends on phase-space coordinates. In this context, the qubit-flip operator in spinor space reads $\sigma^{(P)}_y\otimes\sigma^{(S)}_y = -i\,\gamma^{\2}$, which is applied after the complex conjugate operation,
\begin{equation}
	\widetilde{\gamma^{0}\,\omega^{\pm}_{n,r}} = (-i\gamma^{2})\gamma^{0}\,\omega^{\pm}_{n,r}\,(-i\gamma^{2}),
\end{equation}
not affecting the density matrices in Eqs.~(\ref{9996})-(\ref{9999}) since they have real components.
The promised formula for the quantum concurrence squared is written as \cite{BernardiniEPJP}
\begin{eqnarray}
\int_{_{-\infty}}^{^{+\infty}}\hspace{-.5 cm}ds\,\int_{_{-\infty}}^{^{+\infty}}\hspace{-.5 cm}dk_x \,	\mathcal{C}^2[\omega^{\pm}_{n,r}] &=& \int_{_{-\infty}}^{^{+\infty}}\hspace{-.5 cm}ds\,\int_{_{-\infty}}^{^{+\infty}}\hspace{-.5 cm}dk_x \, Tr[{\gamma^{0}\,\omega}^{\pm}_{n,r}\,\widetilde{\gamma^{0}\,\omega^{\pm}_{n,r}}] \nonumber \\
	&=& 2 (\eta_n B_n ) ^2,
\end{eqnarray}
where these parameters are given by Eq.~\eqref{parameters}, namely, $B_{n} = \frac{\sqrt{2n\,e\mathcal{B}}}{E_{n} +m}$ and $\eta_{n}=\frac{E_{n} +m}{2E_{n}}$, related to the fermion physical parameters.

One notices that while the square root of the above expression is clearly well-defined, the square root of the unintegrated expression cannot be defined locally in phase space, since it depends on the products of functions that take negative values. Thus, one keeps the squared expression as the relevant quantifier for intrinsic entanglement. 

However, a distinct basis is possible since the energy levels are degenerate. This is summarized by the basis change in Eq.~\eqref{superposition}, which introduces the superposition angle $\theta$,
\begin{equation}
\omega_{n,\theta} = \sin^2 (\theta) \, \omega ^+ _{n,1} + \cos^2 (\theta) \, \omega ^- _{n,1} + \sin(\theta) \cos(\theta) \, \Omega_n.
\end{equation}
One finally evaluates the quantum concurrence squared as

\begin{equation}
\int_{_{-\infty}}^{^{+\infty}}\hspace{-.5 cm}ds\,\int_{_{-\infty}}^{^{+\infty}}\hspace{-.5 cm}dk_x \,	\mathcal{C}^2[\omega_{n,\theta}] = 2\eta_n ^2 \, \big(B_n \cos(2\theta) - A_n \sin(2\theta)\big)^2,
\end{equation}
where now $	A_{n}= \frac{k_z}{E_{n} +m}$. Thus, for $\theta = \pi (k+1/2)$, with integer $k$, one recovers the previous result. Each $\theta$ corresponds to a distinct quantum superposition, and $\tan(2\theta) = B_n/A_n$ identifies states that are spin-parity separable for all $A_n \neq 0$ and $\cos(2\theta) \neq 0$.
\section{Phase-space integration of Dirac-Gaussian states}\label{AppC}
The quantum purity in Eq.~\eqref{puritymixed} can be calculated using polar coordinates,
{\footnotesize
	\begin{eqnarray}
		\mathcal{P}_\mathcal{W} &=& 2\pi\int^{+\infty}_{-\infty} \hspace{-.5cm} {dx}\int^{+\infty}_{-\infty} \hspace{-.5cm}{dk_x}\, \Tr[\left( \gamma_0\mathcal{W} (x,\, k_x) \right) ^2] \nonumber \\ 
		&=& \frac{(1-z)^2}{4}\int_{_{0}}^{^{+\infty}}\hspace{-.5 cm}dr \, r \, \Bigg\{\frac{z^2+1}{(z+1)^2} \exp\left[4r^2\left( \frac{z}{z+1} - 1/2 \right)\right] + \exp[-2r^2] \nonumber \\ &+& \frac{2}{z+1} \exp\left[2r^2\left( \frac{z}{z+1} - 1 \right)\right] \Bigg\} \nonumber \\ 
		&=&
		\frac{(z-1)(z^2 -2)}{4(z+1)},
\end{eqnarray}}
where the final result is obtained by noticing that a change of variable $u = r^2$ yields an exponential integrand. 

Eq.~\eqref{entropy1} is obtained using the same method. Explicitly,

\begin{eqnarray}
	\mathcal{I}_{\{x,k_x\}} &=& 1- 4\pi^2 \int_{_{0}}^{^{+\infty}}\hspace{-.5 cm}dr \, r \,\Tr[\mathcal{W} \gamma_0] ^2 \nonumber\\
	&=&	 1- (1-z)^2\int_{_{0}}^{^{+\infty}}\hspace{-.5 cm}dr \, r \, \bigg\{ \exp[-2r^2] + \exp\left[4r^2\left(\frac{z}{z+1} - \frac{1}{2}\right) \right]\nonumber\\
	&+& 2 \exp\left[2r^2\left(\frac{z}{z+1} - 1\right) \right] \bigg\} \nonumber\\
	&=&
	\frac{z}{2}\bigg(2 + z - z^2\bigg).
\end{eqnarray}

Turning to Eq.~\eqref{entropy2}, one now calculates the entropy with respect to the spin-parity Hilbert space (cf. (\ref{entropysp})). The phase-space degrees of freedom are removed by integrating Eqs.~(\ref{mixedGaussian1})-(\ref{mixedGaussian2}),

\begin{eqnarray}
	2\pi \,\int_{_{0}}^{^{+\infty}}\hspace{-.5 cm}dr \, r \,\mathcal{W}_{11} &=& \frac{z}{4}, \\
	2\pi \,\int_{_{0}}^{^{+\infty}}\hspace{-.5 cm}dr \, r \, \mathcal{W}_{22} 	&=& \frac{2-z}{4}.
\end{eqnarray}
Now, in a similar fashion to Eq.~\eqref{entropy1}, the linear entropy is calculated by tracing over the square of the spin-parity density matrix. Then, the corresponding linear entropy (cf. \eqref{entropysp}) can be calculated,
\begin{eqnarray}
	\mathcal{I}^{SP} &=& 1 -4\pi^2\Tr\left[\left(\gamma_{0} \,\int_{_{0}}^{^{+\infty}}\hspace{-.5 cm}dr \, r \,\mathcal{W} \right)^2\right] \nonumber \\
	&=& 1- 2\left[\left(\frac{z}{4}\right)^2 + \left(\frac{2-z}{4}\right)^2 \right] \nonumber \\
	&=& \frac{1}{2}\left(1 + z - \frac{z^2}{2}\right),
\end{eqnarray}
which is the desired result.
 
\section{Dimensionless thermodynamic functions} \label{AppD}
Following a similar procedure described in \cite{Santos} for non-commutative graphene, one extracts other thermodynamics functions from the partition function,
\begin{equation}
	\mathcal{Z}=\sum_{n}\exp(-\beta E_{n}),\quad\beta=\frac{1}{k_{B}T},
\end{equation}
where $k_{B}$ is the Boltzmann constant and $T$ is the equilibrium
temperature. Whereas the partition function is dimensionless, other functions derived from it are not. For instance, $E_n$ is temporarily regarded as a dimensionfull quantity,
\begin{equation}
E_n = \sqrt{(mc^2)^2 + (c \hbar k_z)^ 2 + 2n\,e\mathcal{B} \hbar \, c^ 2}.
\end{equation}
Any term that appears in the energy eigenvalue can be used as a scaling parameter. Choosing the last one, one introduces dimensionless variables as
\begin{equation}
	\tilde{\beta}= \frac{1}{\tilde{T}}, \quad \tilde{T} = \frac{1}{c \sqrt{2e\mathcal{B} \hbar}} \left(\frac{1}{\beta} \right)
\end{equation}
where ``$\sim$'' stands for a dimensionless quantity. This choice yields an analogous scaling to Dirac electrons in graphene, where the speed of light is replaced by the fermi velocity.

Other thermodynamic functions can also be written in terms of these new variables, namely, the free energy $F$, the mean energy $U$, the entropy $S$ and the specific heat $C$. One then obtains dimensionless quantities simply by noticing that 
\begin{equation}
	\beta = \tilde{\beta} \left( \frac{1}{c \sqrt{2e\mathcal{B} \hbar}} \right)
\end{equation}
has units of inverse energy. Thus, the first two can be rescaled as 

\begin{align}
	\tilde{F} &= \frac{F}{c \sqrt{2e\mathcal{B} \hbar}} = -\frac{1}{\tilde{\beta}}\ln \mathcal{Z}, \\
	\tilde{U} &= \frac{U}{c \sqrt{2e\mathcal{B} \hbar}} = - \frac{\partial}{\partial\tilde{\beta}}\ln \mathcal{Z}.
\end{align}
Conversely, the entropy and the specific heat can be simply written as 
\begin{align}
	\tilde{S} &= \frac{S}{k_B} = \tilde{\beta}^{2}\frac{\partial \tilde{F}}{\partial\tilde{\beta}}, \\
	\tilde{C} &= \frac{C}{k_B} = \tilde{\beta}^{2}\frac{\partial \tilde{U}}{\partial\tilde{\beta}} .
\end{align}
Throughout the paper, one uses natural units, with $k_B = c = \hbar = 1$. This is simply equivalent to choosing the dimensionless set of physical observables. To avoid confusion, one adopts the notation $\mu = \tilde{\beta}$ in natural units, so that $\mu$ stands for the inverse temperature normalized by the magnetic field. This parameter will be used for expanding the functions above. In the paper, the ``$\sim$'' notation is omitted. 

\section{Residue of the zeta function} \label{AppE}
The residue theorem is once again applied to $z=1$, by using an alternative representation of the zeta function in Eq.~\eqref{residue}, an analytic continuation in terms of a double sum of binomial coefficients $\binom{n}{m}$ \cite{Hasse}, i.e. 
\begin{eqnarray}
	\Res\big(\zeta(z,\kappa), 1\big) &=&	\lim_{z \rightarrow 1} \, (z-1) \left( \frac{1}{z-1} \sum_{n=0} ^\infty \frac{1}{n+1} \sum_{m=0} ^n (-1)^m \binom{n}{m} (\kappa + m)^{1-z} \right) \nonumber\\
	&=& \sum_{n=0} ^\infty \frac{1}{n+1} \sum_{m=0} ^n (-1)^m \binom{n}{m} \nonumber \\
	&=& 1 + \sum_{n=1} ^\infty \frac{1}{n+1} (1 - 1)^n \nonumber \\
	&=& 1,
\end{eqnarray}
where the binomial theorem has been applied in the third equality.

\section{Finite temperature correlations }
\label{AppF}
In order to calculate each component of the ensemble in Eq.~\eqref{ensemble1}, one recalls that if $\kappa = 0$, the one-particle parameters for excited levels simplify to $B_n = 1, \eta_n = 1/2, A_n = 0$. Thus, for $n \neq 0$, the matrix (\ref{9997}) simplifies to 
\begin{eqnarray}
	\omega^-_{n,1}(s,\,k_x) &=& \frac{1}{2} \left( \begin{array}{cccc} 
		0 & 0& 0& 0\\
		0 & \mathcal{L}_{n} & \mathcal{M}_{n}& 0 \\ 
		0 & -\mathcal{M}_{n} & -\mathcal{L}_{n-1}& 0 \\ 
		0 & 0&0& 0 \\ 
	\end{array} \right).
\end{eqnarray}
However, the ground state has a distinct normalization, depending on the value of $\eta_0$. The simplest choice is $\eta_0 = 1$ ($A_0=0$) since there are less components to evaluate. This choice does not affect the results at absolute zero or in the infinite-temperature limit. The ground state reads

\begin{eqnarray}
	\omega^-_{0,1}(s,\,k_x) &=& \left( \begin{array}{cccc} 
		0 & 0& 0& 0\\
		0 & \mathcal{L}_{0} & 0&0 \\ 
		0 & 0 & 0&0 \\ 
		0 & 0 & 0& 0 \\ 
	\end{array} \right).
\end{eqnarray}

Thus, the phase-space entropy in Eq.~\eqref{ikxthermal} can be calculated, by recalling that the thermalized ensemble has been introduced in Eq.~\eqref{ensemble1}. The result reads
\begin{eqnarray}
	\mathcal{I}_{\{x,k_x\}} &=& 1- \frac{2\pi}{\sqrt{e\mathcal{B}}} \int_{_{-\infty}}^{^{+\infty}}\hspace{-.5 cm}dx\,\int_{_{-\infty}}^{^{+\infty}}\hspace{-.5 cm}dk_x \, \Tr[\mathcal{W}_{TE}\gamma_{0}] ^2 \nonumber \\
	&=& 1- \frac{2\pi}{\mathcal{Z}^2\sqrt{e\mathcal{B}}} \int_{_{-\infty}}^{^{+\infty}}\hspace{-.5 cm}dx\,\int_{_{-\infty}}^{^{+\infty}}\hspace{-.5 cm}dk_x \, \left( \frac{1}{2} \sum_{n=1} p_n \nonumber \mathcal{L}_{n-1} + \frac{1}{2} \sum_{n=1} p_n \mathcal{L}_{n} + p_0\mathcal{L}_0 \right)^2 \\
	&=& 1- \frac{1}{ \mathcal{Z}^2} \left( \frac{1}{2} \sum_{n=1} (p_n) ^2 + \frac{ p_0 ^2}{2} + \frac{1}{2} \sum_{n=1} p_n\, p_{n+1} + \frac{1}{2} p_0\,p_1 \right) \nonumber \\
	&=& 1 - \frac{\mathcal{Z}(2\mu)}{2\mathcal{Z}^2 (\mu) } - \frac{p_0 ^2}{2\mathcal{Z}^2} - \frac{1}{2 \mathcal{Z}^2 } \sum_{n = 0} p_n \, p_{n+1},
\end{eqnarray}
where the phase-space integrals in the second line are performed using the orthogonality of Laguerre polynomials from Eq.~\eqref{norr0w1}. In the last line, the series and the third term become negligible. To see this, one recalls that $\mathcal{Z}^2(\mu) = (\sum p_n)^2$, so that the denominator is always greater than the numerator. Moreover, for high-temperatures, all $p_n$ are arbitrarily small. Thus, the third and fourth terms can be dropped.

The spin-parity entropy in Eq.~\eqref{ispthermal} needs to be considered separately. The expression to be calculated is 

\begin{eqnarray}
	\mathcal{I}^{SP} &=& 1 - \Tr\left[\left(\gamma_{0}\,\int_{_{-\infty}}^{^{+\infty}}\hspace{-.5 cm}dx\,\int_{_{-\infty}}^{^{+\infty}}\hspace{-.5 cm}dk_x \,\mathcal{W}_{TE} (s,\,k_x)\right)^2\right]
\end{eqnarray}
where again
\begin{equation}
	\mathcal{W}_{TE}(s,\,k_x) = \frac{1}{\mathcal{Z}}\sum_{n=0}^{\infty} p_n \left( \omega ^- _{n,1} \right).
\end{equation}
No integration is needed, since the functions $\mathcal{L}_{n}$ are normalized, and $\mathcal{M}_{n}$ vanish upon phase-space averaging (cf. \eqref{norr0w2}).

Now, using the notation $a_{ij}$ to indicate each element of the $4 \times 4$ matrix after integration,

\begin{eqnarray}
	a_{22} &=& \frac{1}{2}\sum_{n=1} p_n + p_0 , \nonumber \\
	a_{33} &=& -\frac{1}{2}\sum_{n=1} p_n.
\end{eqnarray}
Using $\mathcal{Z}= \sum_{n=1} p_n + p_0$,
\begin{eqnarray}
	\mathcal{I}^{SP} &=&1- a_{22} ^2 - a_{33}^2 \nonumber \\
	&=& p_0^2+ \frac{1}{2}(\mathcal{Z}-p_0)^2 + (\mathcal{Z}-p_0)p_0 \nonumber \\ 
	&=& \frac{1}{2}\left( 1 - \frac{p_0^2}{\mathcal{Z}^2}\right),
\end{eqnarray}	
the result shown in Eq.~\eqref{ispthermal}.


\begin{thebibliography}{99}
\bibitem{Andersen2}
U. L. Andersen, J. S. Neergaard-Nielsen, P. van Loock, and A. Furusawa, Nat. Phys. {\bf 11}, 713 (2015).
\bibitem{Vedral}
V. Vedral, Phys. Rev. Lett. {\bf 90}, 050401 (2003).
\bibitem{Henderson}
L. Henderson and V. Vedral, J. Phys. A: Math. Gen. {\bf 34}, 6899 (2001).
\bibitem{n024}
W. K. Wootters, Phys. Rev. Lett. {\bf 80}, 2245 (1998);
W. K. Wootters, Quantum Inf. Comput. {\bf 1}, 27 (2001).	
\bibitem{Weedbrook}
C. Weedbrook, S. Pirandola, R. Garc{\'i}a-Patr{\'o}n, N. J. Cerf, T. C. Ralph, J. H. Shapiro, and S. Lloyd, Rev. Mod. Phys. {\bf 84}, 621 (2012). 
\bibitem{n010}
V. A. S. V. Bittencourt, S. S. Mizrahi and A. E. Bernardini, Ann. Phys. {\bf 355}, 35 (2015).
\bibitem{extfields}
V. A. S. V. Bittencourt and A. E. Bernardini, Ann. Phys. {\bf 364}, 182 (2016).
\bibitem{BernardiniEPJP}
A. E. Bernardini, The Eur. Phys. J. Plus {\bf 135}, 675 (2020).
\bibitem{n001}
L. Lamata, J. Le\'{o}n, T. Sch\"{a}tz and E. Solano, Phys. Rev. Lett. {\bf 98}, 253005 (2007).
\bibitem{n002}
J. Casanova, J. J. Garc\'{i}a-Ripoll, R. Gerritsma, C. F. Roos and E. Solano, Phys. Rev. A {\bf 82}, 020101(R) (2010).
\bibitem{n004}
A. Bermudez, M. A. Martin-Delgado and E. Solano, Phys. Rev. A {\bf 76}, 041801(R) (2007).
\bibitem{n005}
T. G. Tenev, P. A. Ivanov and N. V. Vitanov, Phys. Rev. A {\bf 87}, 022103 (2013).
\bibitem{n006}
L. Lamata, J. Casanova, R. Gerritsma, C. F. Roos, J. J. Garc\'{i}a-Ripoll and E. Solano, New Journal of Physics {\bf 13}, 095003 (2011).
\bibitem{graph03}
E. McCann and M. Koshino, Rep. Prog. Phys. {\bf 76}, 056503 (2013).
\bibitem{graph04}
A. V. Rozhkov, A. O. Sboychakov, A. L. Rakhmanov and F. Nori, Phys. Rep. {\bf 648}, 1-104 (2016). 
\bibitem{CastroNeto}
A. H. Castro Neto, F. Guinea, N. M. R. Peres, K. S. Novoselov and A. K. Geim {\bf 81}, 109 (2009).
\bibitem{MeuPRB}
V. A. S. V. Bittencourt and A. E. Bernardini, Phys. Rev. B {\bf 95}, 195145 (2017).
\bibitem{PRBPRB18}
V. A. S. V. Bittencourt, A. E. Bernardini and M. Blasone, Phys. Rev. B {\bf 97}, 125435 (2018).
\bibitem{New}
V. A. S. V. Bittencourt and A. E. Bernardini, J. Phys. B.: At. Mol. Opt. Phys. {\bf 50}, 075501 (2017).
\bibitem{MeuPRA}
V. A. S. V. Bittencourt, A. E. Bernardini and M. Blasone, Phys. Rev. A {\bf 93}, 053823 (2016). 
\bibitem{Thaller}
B. Thaller, {\it The Dirac Equation}, (Springer-Verlag, 1992).
\bibitem{Bermudez}
A. Bermudez, M. A. Martin-Delgado and E. Solano, Phys. Rev. Lett. {\bf 99}, 123602 (2007).
\bibitem{Rusin}
T. M. Rusin and W. Zawadski, Phys. Rev. D, {\bf 82}, 125031 (2010).
\bibitem{Greiner}
W. Greiner, {\em Relativistic quantum mechanics}, (Springer, 2000).
\bibitem{Bera}
A. Bera, T. Das, D. Sadhukhan, S. S. Roy, A. Sen(De), and U. Sen, Rep Prog. Phys. {\bf 81}, 024001 (2018). 
\bibitem{Olivier}
H. Olivier and W. H. Zurek, Phys. Rev. Lett. {\bf 88}, 017901 (2001). 
\bibitem{Luo}
S. Luo Phys. Rev. A {\bf 77}, 042303 (2008).
\bibitem{PRA2023}
C. F. Silva and A. E. Bernardini, Phys. Rev. A, {\bf 107}, 042220 (2023).
\bibitem{1986}
H. T. Elze, M. Gyulassy and D. Vasak, Nucl. Phys. B {\bf 276}, 706 (1986).
\bibitem{1987}
D. Vasak, M. Gyulassy and H. T. Elze, Ann. Phys. {\bf 173}, 462 (1987). 
\bibitem{Wigner}
E. Wigner, Phys. Rev. {\bf 40}, 749 (1932).
\bibitem{Weickgenannt}
N. Weickgenannt, X. L. Sheng, E. Sperenza, Q. Wang and D. H. Rischke, Phys. Rev. D {\bf 100}, 056018 (2019).
\bibitem{Zhuang}
P. Zhuang and U. Heinz, Ann. Phys. {\bf 245}, 311-338 (1996).
\bibitem{Case}
W. B. Case, Am. J. Phys. {\bf 76}, 937-946 (2008).
\bibitem{PRA2021}
C. F. Silva and A. E. Bernardini, Phys. Rev. A {\bf 104}, 052213 (2021).
\bibitem{Andersen}
U. L. Andersen, G. Leuchs, and C. Silberhorn, Laser Photonics Rev. {\bf 4}, 337 (2010).
\bibitem{BernardiniEPJP2}
A. E. Bernardini, V. A. S. V. Bittencourt and M. Blasone, The Eur. Phys. J. Plus {\bf 135}, 320 (2020).
\bibitem{Gibbs}
J. W. Gibbs, {\em Elementary principles in statistical mechanics: developed with especial reference to the rational foundations of thermodynamics}, (C. Scribner's sons, 1902).
\bibitem{Houca}
R. Hou\c{c}a and A. Jellal, Phys. Scr., {\bf 94}, 105707 (2019). 
\bibitem{Artin}
E. Artin, {\em The gamma function}, (Courier Dover Publications, 2015).
\bibitem{Hasse}
H. Hasse, Math. Z., {\bf 32}, 458 (1930).
\bibitem{Santos}
V. Santos, R.V. Maluf, and C.A.S. Almeida, Ann. Phys. {\bf 349}, 402 (2014).
\bibitem{BernardiniPHYSA}
A. E. Bernardini, Phys. A {\bf 557}, 124889 (2020).
\bibitem{Goerbig}
M. O. Goerbig, Rev. Mod. Phys. {\bf 83}, 1193 (2011).
\bibitem{Kumar}
P. S. Kumar, I. F. Herbut and R. Ganesh, Phys. Rev. Research {\bf 2}, 033035 (2020).
\bibitem{Sun}
J. Sun, H. Guo and S. Feng, Phys. Rev. Research {\bf 3}, 043223 (2021).

\end{thebibliography}
\end{document}